\documentclass[a4paper,11pt]{article}
\usepackage{pos}
\usepackage{xspace} 
\usepackage{hyperref}
\usepackage{url} 
\usepackage{multicol}
\usepackage{wrapfig}
\newcommand{\vareps}{\varepsilon}
\newcommand{\Fermi}{\emph{Fermi}\xspace}

\newcommand{\txs}{TXS\,0506+056\xspace} 
\newcommand{\pks}{PKS\,1502+106\xspace}
\newcommand{\pksb}{PKS\,B1424-418\xspace}
\newcommand{\hsp}{3HSP\,J095507.9+355101\xspace} 
\newcommand{\ictxs}{IC-170922A\xspace}
\newcommand{\ichsp}{IC-200107A\xspace}
\newcommand{\icpks}{IC-190730A\xspace}

\newcommand{\gRay}{$\gamma$-ray\xspace}
\newcommand{\gRays}{$\gamma$-rays\xspace}

\newcommand{\be}{\begin{equation}}
\newcommand{\ee}{\end{equation}}
\newcommand{\ba}{\begin{eqnarray}}
\newcommand{\ea}{\end{eqnarray}}
\newcommand{\bi}{\begin{itemize}}
\newcommand{\ei}{\end{itemize}}

\title{High-Energy Neutrino Emission from Blazars}
 \ShortTitle{High-Energy Neutrino Emission from Blazars}

\author*[a]{Foteini Oikonomou}

\affiliation[a]{
Institutt for Fysikk, Norwegian University of Science and Technology, Trondheim, Norway}

\emailAdd{foteini.oikonomou@ntnu.no}

\abstract{Active galactic nuclei (AGN) with relativistic jets are the most
powerful persistent astrophysical sources of electromagnetic radiation
in the Universe. Blazars are the most extreme subclass of AGN with
jets directed along the line of sight of the observer. 
Their high-energy photon emission dominates
the extragalactic gamma-ray sky and reaches multi-TeV energies. This
demonstrates that blazars accelerate particles to very high energies. It
has long been suspected that blazars may also accelerate protons to
very high energies and thus be cosmic neutrino sources. Being
extremely rare objects in addition to being bright, blazars are among
the most readily testable neutrino candidate source classes.
Several multi-messenger monitoring campaigns have recently been
triggered in response to high-energy neutrinos observed with the
IceCube Neutrino Observatory from the direction of blazars. In this
contribution, I summarise the theoretical interpretation of these
observations and give an overview of the possible role of blazars as
neutrino sources in light of the experimental results. 
}

\FullConference{37$^{\rm{th}}$ International Cosmic Ray Conference (ICRC 2021)\\
		July 12th -- 23rd, 2021\\
		Online -- Berlin, Germany}

\begin{document}
\maketitle

\section{Introduction}

\noindent The discovery of high-energy astrophysical neutrinos was reported in 2012 by the IceCube Neutrino Observatory~\cite{Aartsen:2013bka}. Since then, the field of neutrino astronomy has made considerable progress, with the detection of neutrinos in multiple analysis channels~\cite{IceCube:2020acn,IceCube:2021uhz}, and with the first indication of a joint source of \gRays and neutrinos~\cite{IceCube:2018dnn}. In tandem with these experimental breakthroughs, significant theoretical efforts are being made to determine the origin of astrophysical neutrinos. The proposed neutrino source classes include Galactic sources, \gRay bursts, jetted and non-jetted Active Galactic Nuclei as well as calorimetric environments such as starburst galaxies to name some of the possibilities~\cite{Stecker:1978ah,Berezinsky:1992wr,PhysRevLett.66.2697,Meszaros:2001ms,Loeb:2006tw}. The isotropy of the neutrino arrival direction distribution rules out a purely Galactic origin and implies that the majority of the neutrinos have an extragalactic origin. 

A particularly promising source candidate are blazars. These are AGN with jets in our line of sight. Blazars are very rare objects. Only a small fraction of galaxies host an AGN with a relativistic jet, and only $\sim1/100$ of them point in our line of sight. Thus, the number density of blazars inferred with the \Fermi Large Area Telescope is approximately $\lesssim 10^{-6}~{\rm Mpc}^{-3}$~\cite{Ajello:2013lka}. The apparent luminosity of a powerful blazar can be of order $10^{49}~{\rm erg/s}$, $10^7$ times larger than the luminosity of the Milky Way. Here the luminosity is the apparent luminosity, which benefits from a relativistic boost in our direction. A similar boost is experienced for neutrinos if they are produced in blazar jets. 

Blazars are the dominant sources of GeV and TeV \gRays. This means that as a population they are excellent high-energy accelerators and that they comfortably accelerate particles to above 100~TeV. It is reasonable to assume that some of the accelerated particles are protons, in which case, if interactions with ambient matter or photons occur, high-energy neutrinos will be produced. Thus, blazars have been proposed since the early nineties as ideal sources of neutrinos~\cite{1992A&A...260L...1M,1993A&A...269...67M,Atoyan:2001ey,Muecke:2002bi,2003ApJ...586...79A,Neronov:2002xv,Dermer:2007me, 2014PhRvD..90b3007M,Tavecchio:2014xha,Petropoulou:2015upa,Petropoulou:2016ujj,Padovani:2015mba,Gao:2016uld,Rodrigues:2017fmu,Palladino:2018lov,Rodrigues:2020pli}. 

This contribution summarises the key points of the highlight talk I gave at the ICRC 2021. In Section~\ref{sec:production}, a brief overview of neutrino production mechanisms in blazars is given. In Section~\ref{sec:diffuse} the current state of knowledge of the possible contribution of blazars to the diffuse neutrino flux measured with IceCube is reviewed. In Section~\ref{sec:point_source}, the capabilities of blazars to be detected as neutrino point sources is reviewed, with a focus on three recently observed neutrino-blazar associations for which detailed modelling was performed. A discussion and outlook follow in Section~\ref{sec:discussion}. 

\section{Neutrino production in blazars} 
\label{sec:production} 

\begin{wrapfigure}{l}{0.42\textwidth}
\includegraphics[trim=0cm 9cm 44.5cm 4.5cm, clip, width = 0.41\textwidth]{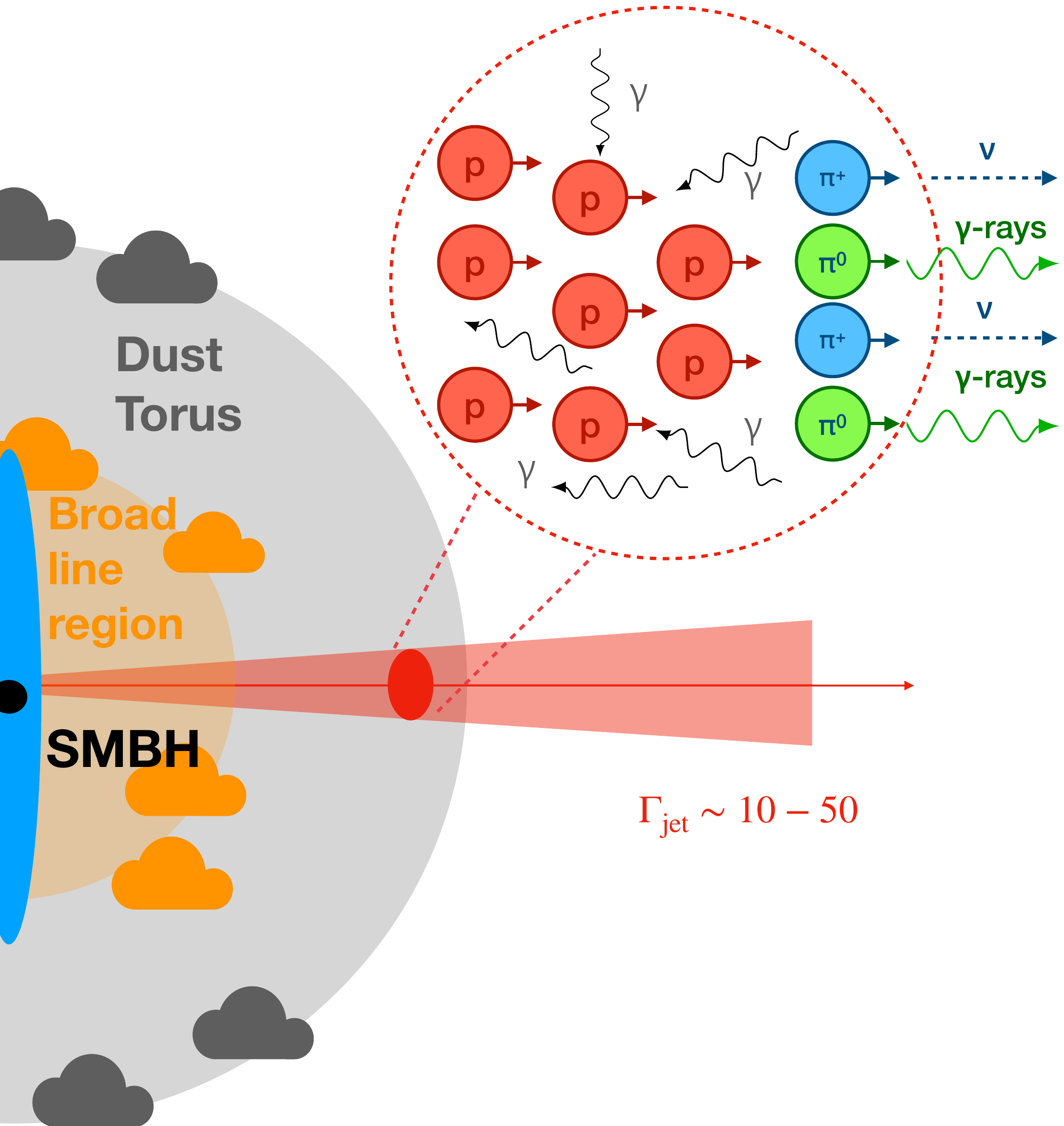}
\vspace{-0.2cm}
\caption{Schematic representation of neutrino production in a blazar jet through $p\gamma$ interactions. High-energy protons present in the emitting region interact with photons to produce charged and neutral pions leading to comparable fluxes of neutrinos and \gRays.
 \label{fig:sketch}\vspace{-0.33cm}}
\end{wrapfigure}

\noindent Neutrinos are produced in photopion ($p\gamma, p + \gamma \rightarrow \pi^+ + n$) or hadronic ($pp, p + p \rightarrow X + N_{\pi} \pi^{\pm}$) interactions of protons and nuclei. Here $N_{\pi}$ is the pion multiplicity. Jetted AGN possess strong radiation fields, particularly at distances close to the base of the jet, associated with emission from the accretion disk surrounding the supermassive black hole (SMBH), but also within the jet itself. Hence, photopion interactions are likely if protons are accelerated by blazars. Some possibilities exist for $pp$ interactions, for example, if dust clouds or a star intersect with accelerated protons, but in general, the $p\gamma$ process is thought to be the prevalent process. An illustration of this, often assumed, scenario is shown in Fig.\,\ref{fig:sketch}. Here, a spherical emitting region is comoving with the jet which has bulk Lorentz factor $\Gamma_{\rm jet}$ which is typically inferred to be in the range $\sim10-50$. Usually, four target photon fields are considered; jet radiation, the radiation from the accretion disk, intercepted accretion-disk radiation reradiated in the ultraviolet range by fast-moving gas clouds (known as the broad-line region), and, further away from the jet base, intercepted accretion-disk radiation reradiated by the dust torus at infrared wavelengths. 

Of the energy lost by protons with energy $\varepsilon_p$ in $p\gamma$ interactions, about 3/8ths go to neutrinos, resulting in the production of neutrinos with all-flavour luminosity, $\varepsilon_{\nu} L_{\varepsilon_{\nu}}\,=\,3/8f_{p\gamma}\,\varepsilon_p\,L_{\varepsilon_p}.$ Here, $f_{p\gamma}$ is the photopion production efficiency, which depends on the number density of available photon targets and the distance over which the interactions take place and $\varepsilon_{\nu/p} L_{\varepsilon_{\nu/p}}$ is the luminosity per logarithmic energy of neutrinos and protons respectively. Each neutrino is produced with energy $\varepsilon_{\nu} \sim0.05 \varepsilon_p$.  The remaining 5/8ths of energy lost by protons result in the production of electrons and \gRays, which generally leads to an electromagnetic cascade inside the accelerator which re-emerges at keV-GeV wavelengths~\cite{Murase:2018iyl}. Thus, neutrino production is accompanied by X-ray and \gRay emission with comparable luminosity. 

The proton luminosity, $L_p\,=\,\int_{\vareps_p}\,L_{\vareps_p}\,d\vareps_p,$ is a key unknown quantity for neutrino model predictions. Upper bounds on it can be derived from electromagnetic (EM), neutrino, and ultra-high energy cosmic ray (UHECR) observations and will be discussed below. The exact value of $f_{p\gamma}$ is also unknown but bounds on this quantity are tighter since in general, it can be determined from the shape of the \gRay spectrum whether \gRays escape the blazar without suffering severe internal attenuation on the same photon fields that facilitate $p\gamma$ interactions~\cite{Waxman:1998yy,Murase:2015xka,Murase:2018iyl}. 

If blazars facilitate intense neutrino production, the EM cascade radiation is an additional spectral signature~\cite{petro_2015}. Alternatively, the absence of EM cascade radiation can be used to set limits on the proton luminosity and thus on the neutrino output of the sources. As was demonstrated in the case of \txs, the absence of EM cascade radiation provides a stronger constraint on the high-energy proton content of the blazar jet than the detection of a single high-energy neutrino~\cite{Ahnen:2018mvi,Cerruti:2018tmc,Gao:2018mnu,Keivani:2018rnh}. 

\begin{figure}
\includegraphics[trim=0.2cm 0.8cm 0.5cm 0.5cm, clip, width = \textwidth]{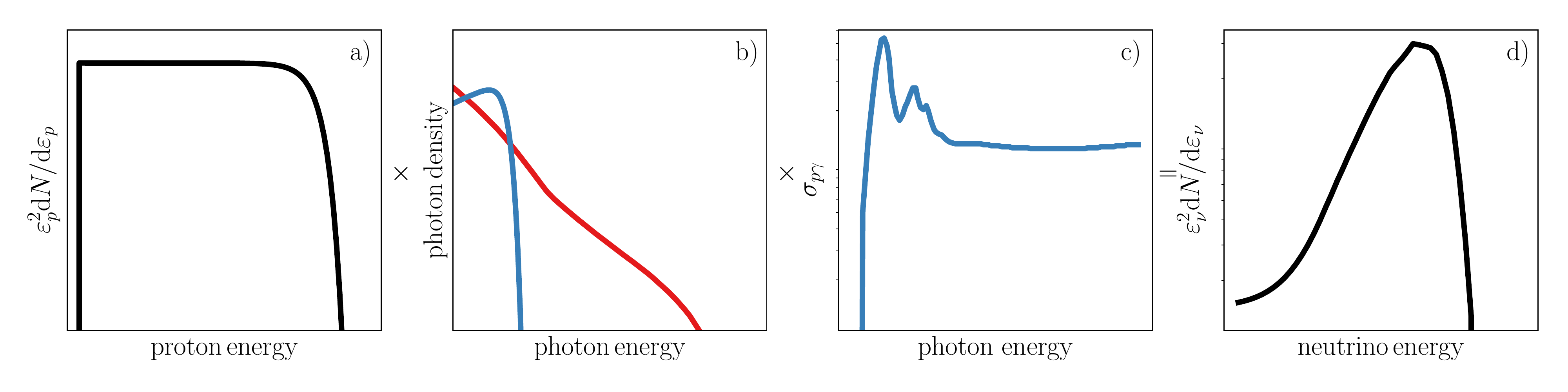}
\vspace{-0.4cm}
\caption{Schematic overview of the expected neutrino spectrum of blazars from $p\gamma$ interactions.
 A population of protons with a spectrum that evolves as ${\rm d}N/{\rm d}\varepsilon_p = \varepsilon^{-2} \exp{(-\varepsilon_p/\varepsilon_{p, \rm max})}$ (panel a)
  interacts with the photons from the blazar jet (red) and possibly stationary photons 
  with a thermal spectrum (blue) in addition (panel b). 
  Such thermal photon fields exist close to the base of many AGN jets and are reprocessed
   radiation from the accretion disk. Due to the resonant shape of the $p\gamma$ 
   cross-section (panel c) and the decreasing number of photons as a function of energy, 
   higher energy protons have access to a larger number of photon targets 
   giving rise to a strongly peaked neutrino spectrum (panel d). All the axes are in double-logarithmic scale. \label{fig:schematic_neutrino_spectrum}}
\end{figure}

Fig.\,\ref{fig:schematic_neutrino_spectrum} gives a schematic summary of the neutrino spectrum produced in photopion interactions of a proton population with a distribution that follows ${\rm d}N/{\rm d} \varepsilon_p \propto \varepsilon_p^{-2}$, with photon fields typical in a blazar. The result is a highly peaked neutrino spectrum in terms of energy flux, even though the proton spectrum is flat. This is due to the resonant shape of the $p\gamma$ cross-section and the decreasing number of photons as a function of energy which is typical in blazars. 

If protons of sufficiently high-energy are present in the jet of a blazar at redshift $z$, the typical energy of the emerging neutrinos is 
$\vareps_{\nu} \approx 1~{\rm PeV}\left(40~{\rm eV}/\vareps_t\right)(1+z)^2$,
for interactions with photons with energy $\vareps_t$ which are stationary with respect to the SMBH. 
Quoted energies are in the observer's frame. For target photons comoving with a jet with bulk Lorentz factor $\Gamma$, the emerging neutrino energy is 
$\vareps_{\nu} \approx 100~{\rm PeV} \left(40~{\rm eV}~/\vareps_t\right)\left(\Gamma/10\right)^2(1+z)^2$.

\section{Blazars as sources of the diffuse IceCube neutrino flux}
\label{sec:diffuse} 

\begin{figure}
\begin{center}
\includegraphics[trim=0.2cm 0.5cm 0.5cm 0.5cm, clip, width = 0.8\textwidth]{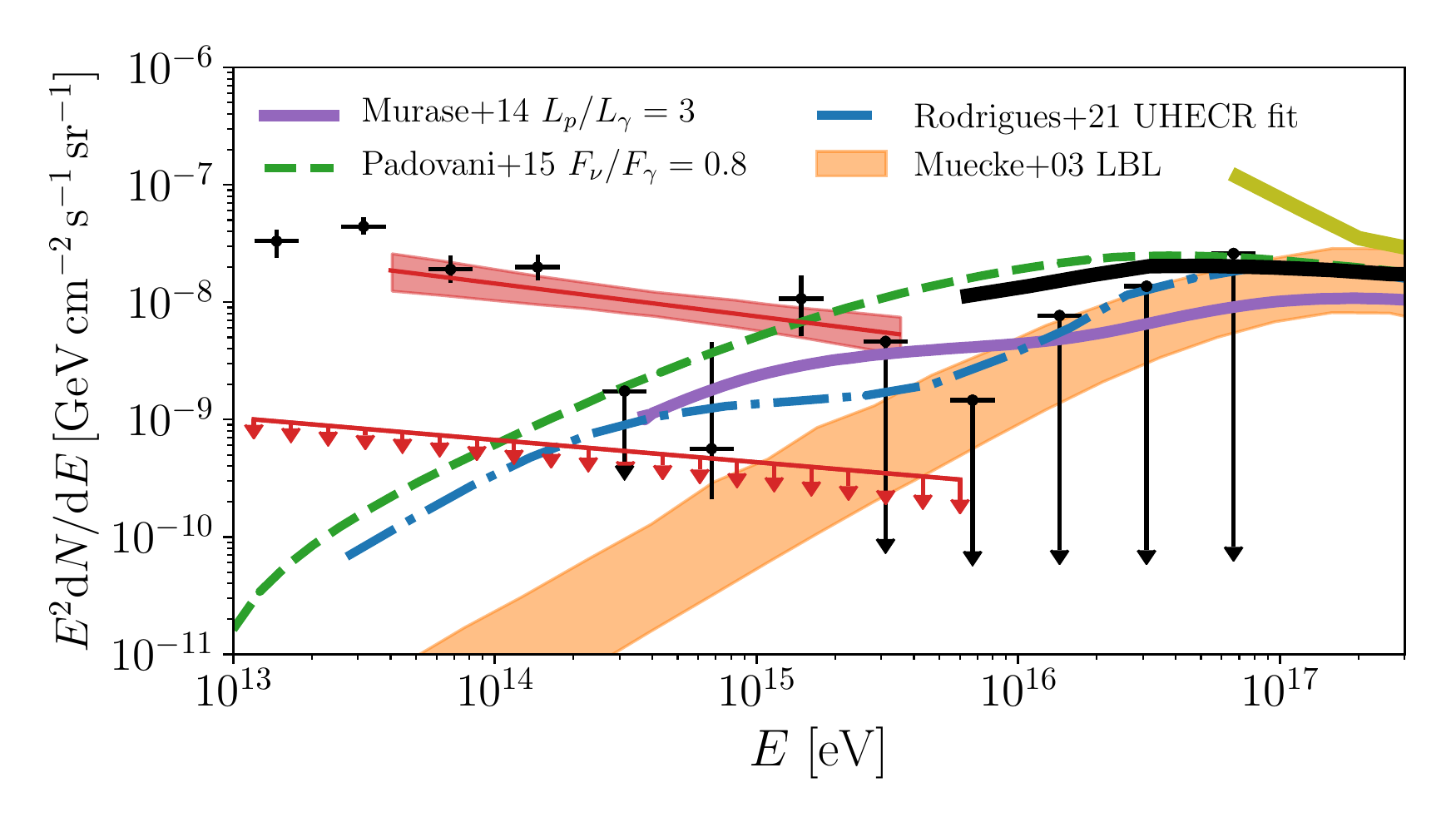}
\vspace{-0.3cm}
\caption{Measured neutrino flux, and predicted diffuse flux of high-energy neutrinos in different blazar models. The red area corresponds to the best-fit and $1\sigma$ uncertainty region of the through-going muon neutrino flux measured with IceCube from~\cite{2019ICRC...36.1017S}. The black data points correspond to the high-energy-starting-event energy spectrum measured with IceCube~\cite{IceCube:2020acn}. The black solid line gives the $90\%$ CL upper limit obtained with nine years of data with the IceCube extremely-high-energy analysis~\cite{Aartsen:2018vtx}. The yellow line gives the $90\%$ CL upper limit on the diffuse neutrino flux from the Pierre Auger Observatory~\cite{Zas:2017Yb}. The red solid line shows the $90\%$ CL upper limit on the average neutrino flux of all blazars from the stacking analysis of~\cite{Huber:2019lrm}. The blazar neutrino models shown correspond to the expected diffuse flux  from~\cite{Muecke:2002bi} (orange band), all-blazar model normalised to the diffuse UHECR flux (corresponding to baryon loading factor $\xi = 3$)~\cite{2014PhRvD..90b3007M} (purple solid line), all-blazar model of~\cite{Padovani:2015mba} (green dashed line), and all-blazar model from a fit to the UHECR spectrum and composition observables of~\cite{Rodrigues:2020pli} (blue dot-dashed line).\label{fig:diffuse}} 
\end{center}
\end{figure}

\noindent Figure~\ref{fig:diffuse} shows the measured neutrino flux and the predicted diffuse flux of high-energy neutrinos in different blazar models. Although the models were developed during a period of almost 20 years and are based on different assumptions, the overall shape and peak energy of the expected neutrino flux among them are comparable. All the models predict that the blazar neutrino flux peaks at $>10$~PeV for the reasons explained above concerning the neutrino production mechanism. This is well beyond the energy range in which the bulk of IceCube neutrinos have been observed, and the observed spectral index of the diffuse astrophysical neutrino flux is significantly softer than the blazar models predict. At the time of writing, the best-fit spectral index is $-2.37\pm0.09$ for the through-going muon neutrino flux, and even softer for the high-energy-starting-event and cascade samples~\cite{IceCube:2020acn,IceCube:2020wum,IceCube:2021uhz}. 

The normalisation of the neutrino flux in the majority of the models is a semi-free parameter, and it is proportional to $L_p$. It is constrained by the non-observation of $>10$~PeV neutrinos with IceCube and the Pierre Auger Observatory (Auger). The baryon loading factor, defined as $\xi = L_p/L_{\gamma}$, 
where $L_{\gamma}$ is the \gRay luminosity of the blazar, is often used to characterise the proton content of the jet. The absence of $>$10~PeV neutrinos thus far, in combination with the model of~\cite{2014PhRvD..90b3007M}, implies that the proton content of blazar jets on average must be $ \xi < 50$. Different model assumptions lead to slightly different limits on $\xi$. 

The upper limits on the all-blazar neutrino flux obtained with IceCube and Auger also constrain the origin of \gRays in the blazar population. For example, in the model of~\cite{Padovani:2015mba} the neutrino flux is assumed to be proportional to the \gRay flux and it is parametrised by the quantity $F_{\nu}/F_{\gamma}$ which is the ratio of the neutrino to \gRay flux.  The current upper limit of the $F_{\nu}/F_{\gamma}$ ratio of 0.08~\cite{IceCube:2021uhz}, implies that at most a few percent of all detected blazar \gRay emission is produced in photohadronic interactions. 

Figure~\ref{fig:diffuse} also shows the $90\%$ CL upper limit on the contribution of \gRay selected (from the 3FHL catalogue~\cite{TheFermi-LAT:2017pvy}) blazars to the IceCube neutrino flux for an assumed $E^{-2.19}$ power-law spectrum, based on the stacking analysis of~~\cite{Huber:2019lrm}. The limit depends on the assumed spectral index of the neutrinos but constrains the relative contribution of the blazars to $<16.7\%$ of the total IceCube flux. As will be summarised in Section~\ref{sec:discussion}, stacking analyses, through which it is investigated whether there is an excess of neutrinos from the directions of the ensemble of resolved sources from a particular population, are powerful and have already strongly constrained several promising candidate source classes as the dominant sources of high-energy astrophysical neutrinos in addition to blazars. Independent upper limits on the relative contribution of blazars to the diffuse neutrino flux have been derived from the absence of clustering and of neutrino point sources in the IceCube data~\cite{Murase:2016gly,Ando:2017xcb,Yuan:2019ucv,Capel:2020txc}, and also disfavour blazars as dominant sources of the IceCube neutrinos. 

The limits on blazars as dominant sources of the IceCube neutrino flux are severe but can be avoided if these blazars which are unresolved in \gRays provide overwhelmingly more neutrinos than the brighter, observed ones, with factors of $\xi = 10^7$ required for the faintest blazars during their longterm emission~\cite{Palladino:2018lov}. However, for well-studied blazars, such values of $\xi$ are ruled out based on the absence of a corresponding spectral signature in X-ray and \gRays. In addition, such a model demands that CR acceleration stops at low energies. As a result, blazars cannot contribute to the diffuse UHECRs in a scenario that avoids the IceCube stacking limits.  

It has been pointed out that the observed high-energy neutrinos, UHECRs and \gRays indicate comparable luminosity densities. This apparent coincidence has been discussed as a possible indication of a joint origin of the three messengers~\cite{Waxman:2013zda,Ahlers:2018fkn,Murase:2018utn}. However, the majority of \gRays originate in blazars~\cite{TheFermi-LAT:2015ykq,Lisanti:2016jub,Zechlin:2016pme} and analyses of IceCube data all but exclude a dominant contribution of blazars to the neutrino flux below 100~TeV. Therefore, a common origin of the bulk of all the three messengers seems unlikely.  

\begin{figure}
\begin{center}
\includegraphics[trim=0.2cm 0.5cm 0.5cm 0.5cm,width = 0.8\textwidth]{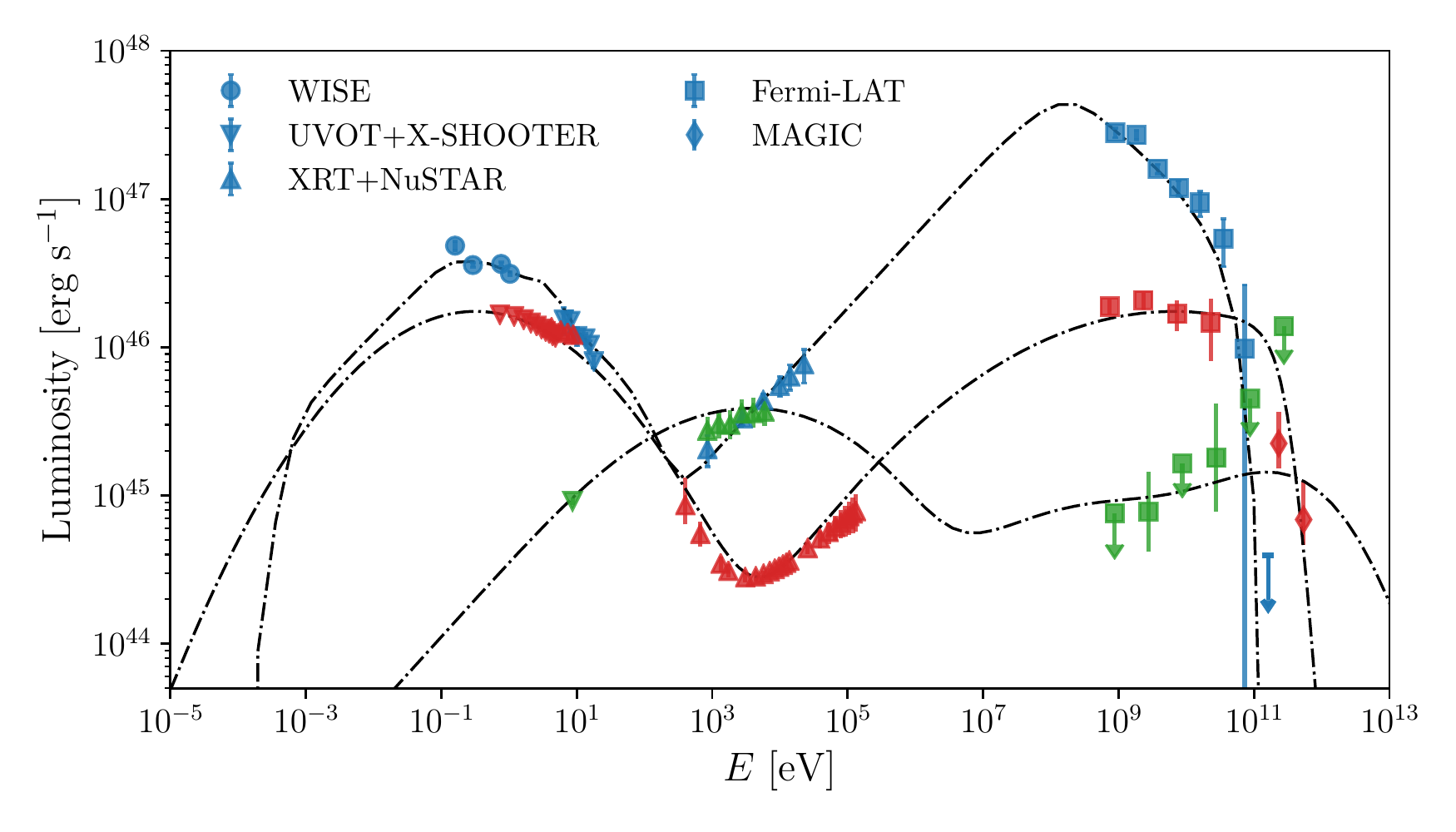}
\vspace{-0.3cm}
\caption{The spectral energy distributions of the blazars \txs (red), \pks (blue), and \hsp(green) at the time of arrival of \ictxs , \icpks , and \ichsp respectively, plotted in terms of the intrinsic source luminosity. 
The model fits are from Refs.~\cite{Oikonomou:2019djc, Oikonomou2021, Petropoulou:2020pqh}. Detailed references to the observations can be found therein.}
\vspace{-0.5cm}
\label{fig:seds}
\end{center}
\end{figure}

\section{Blazars as powerful neutrino point-sources} 
\label{sec:point_source} 

\begin{figure}
\includegraphics[trim=0.1cm 0.0cm 0.45cm 0.4cm, clip, width = 0.653\textwidth]{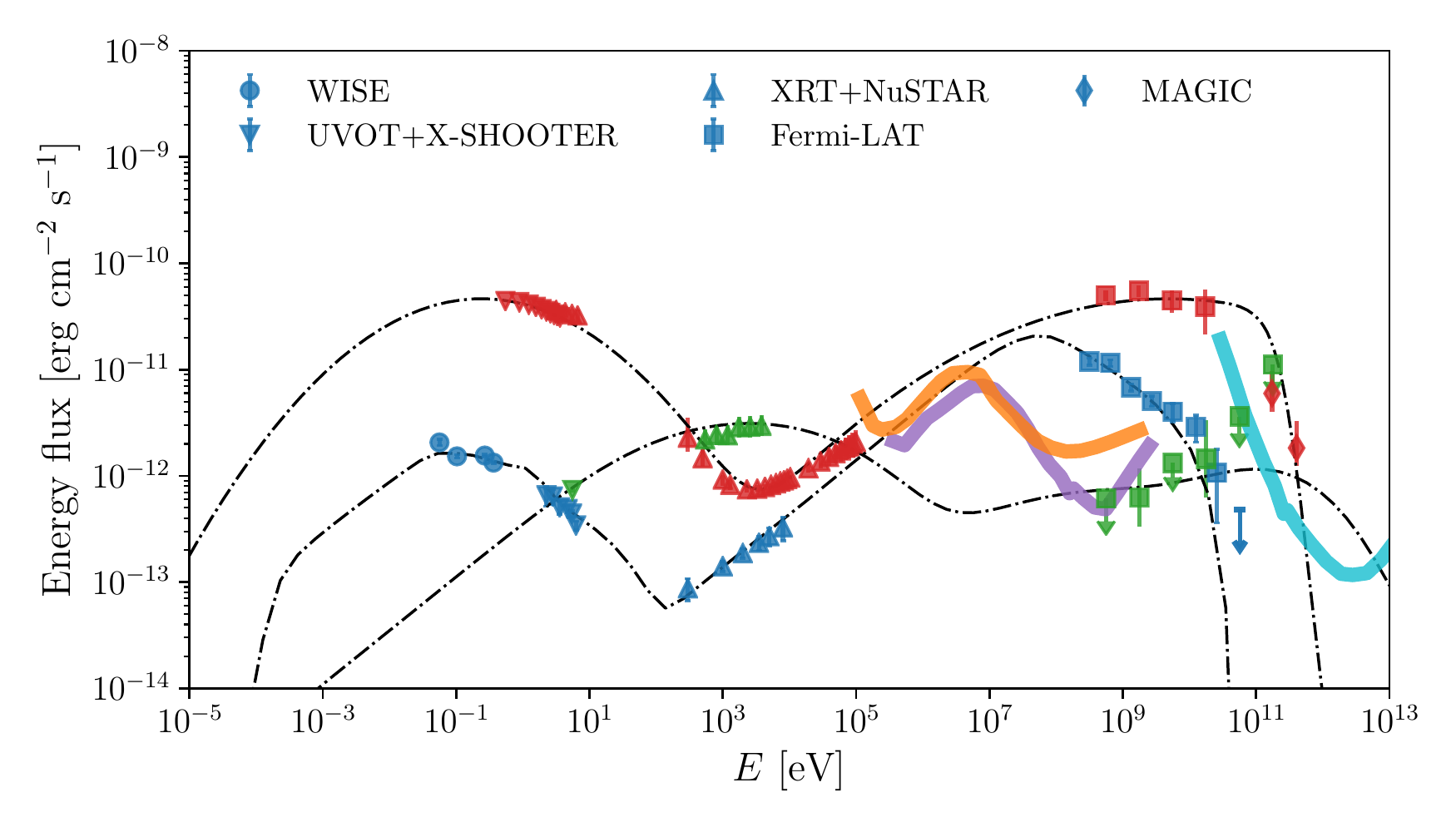}
\includegraphics[trim=0.1cm 0.0cm 0.45cm 0.5cm, clip, width = 0.347\textwidth]{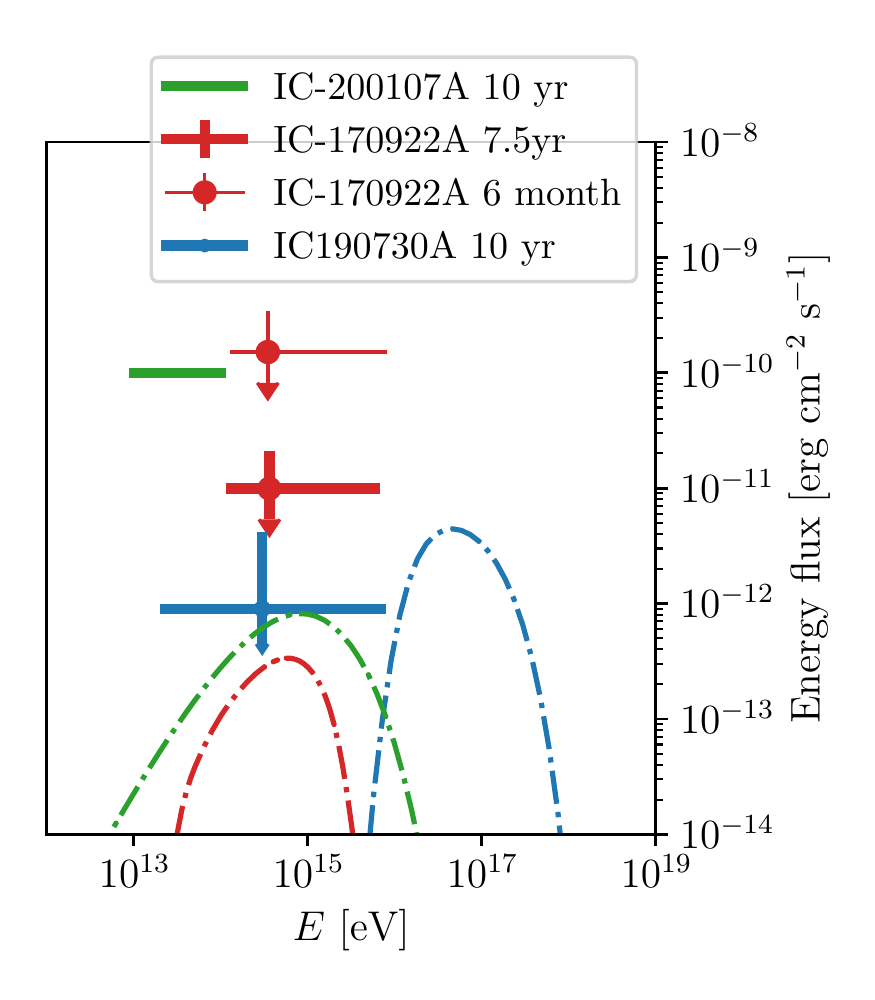}
\vspace{-0.3cm}
\caption{{\bf Left:} Same as the Fig.\,\ref{fig:seds} but shown in terms of the arriving energy flux. The orange line indicates the planned one-year sensitivity of eAstrogram in survey mode~\cite{e-ASTROGAM:2016bph}. The purple line indicates the planned three-year sensitivity of AMEGO~\cite{AMEGO:2019gny}. The blue curve indicates the planned sensitivity of CTA North after 50 hours of observation\cite{cta}. 
{\bf Right:} Model-predicted neutrino energy flux for \txs (red dot-dashed), \pks (blue dot-dashed), and \hsp(green dot-dashed) at the time of arrival of of \ictxs , \icpks , and \ichsp respectively. The neutrino fluxes shown correspond to the most optimistic models of Refs.~\cite{Oikonomou:2019djc,Oikonomou2021,Petropoulou:2020pqh}. Red, blue, and green arrows indicate the $90\%$ CL upper limit on the neutrino flux derived from the detection of \ictxs, \icpks and \ichsp  under the assumption of constant neutrino emission for six months (seven and a half years) - thin (thick) lines for \ictxs, and ten years in the case of \ichsp and \icpks. \label{fig:neutrinos}}
\vspace{-0.3cm}
\end{figure}

\noindent Even though the majority of experimental results suggest that blazars are not the dominant sources of the neutrinos that IceCube has measured thus far, being very rare and extremely bright, blazars may still be detectable neutrino point sources. Some indications of correlations of blazars with high-energy neutrinos have been reported in~\cite{Plavin:2020emb,Giommi:2020hbx,Hovatta:2020lor}, (see also \cite{Franckowiak:2020qrq}). If corroborated by future observations, these results would suggest that a fraction of the neutrino flux is produced by blazars. 

In 2017, \txs was the first flaring blazar to be associated with the high-energy neutrino \ictxs~\cite{IceCube:2018dnn}. It was also the first astrophysical source to be associated with a high-energy neutrino at the $3\sigma$ level. Evidence for additional neutrinos was found in archival IceCube data~\cite{IceCube:2018cha}, the latter constitute $3.5\sigma$ departure from background expectations. During the archival neutrino flare, \txs was not in a flaring \gRay state. 

Periods of flaring EM emission are promising as times of neutrino production. Qualitatively, during the flare, the target material (photons) is more abundant, and, likely, protons are also more abundant during such periods. Thus a non-linear increase of the neutrino flux is predicted in most theoretical models, e.g.~\cite{2014PhRvD..90b3007M,Tavecchio:2014xha,Gao:2016uld}. Model predictions vary considerably for different blazars, and different models lead to expected neutrino counts in IceCube in the range $10^{-5} - $ few neutrinos per source~e.g.\cite{Oikonomou:2019djc,Stathopoulos:2021mli} though models that predict more than a few neutrinos per source are already constrained by the absence of such signal in IceCube. An additional advantage of focusing on flaring emission is that the experimental search is more sensitive when the neutrino signal is transient.  

Since the observation of \ictxs, more blazars have been found in the error circles of high-energy alert neutrinos, albeit with lower significance individually. See e.g.~\cite{Garrappa:2021ihz} in these proceedings, as well as~\cite{Acciari:2021YA,Abbasi:2021XE,Aublin:2021q6,Illuminati:202198,Kadler:20212/}. In~\cite{Oikonomou:2019djc,Oikonomou2021,Petropoulou:2020pqh}, we have performed detailed leptohadronic modelling for three of these sources, in order to study the observed associations from the theoretical point of view. Fig.\,\ref{fig:seds} shows the three studied blazars in terms of their intrinsic luminosities. \pks stands out in terms of its intrinsic \gRay luminosity. \hsp stands out as it has a synchrotron peak at very high energy. Due to this observation, it is classified as an ``extreme'' blazar. In terms of their spectral energy distributions, the three sources are heterogeneous. Fig.\,\ref{fig:neutrinos} left, shows the spectral energy distributions in terms of the arriving energy flux at Earth. \txs had a much higher \gRay flux at the time of the neutrino arrival than the other two sources. This explains the higher statistical significance of this association with respect to the other two. 

\subsection{\txs} 

\noindent \ictxs was detected with energy $\sim290$~TeV. At the time of its arrival, \txs, which lies inside the $50\%$ containment contour of the arrival direction of \ictxs, was undergoing its most extreme recorded \gRay flare. In~\cite{Padovani:2019xcv}, it was demonstrated that \txs is a misclassified BL Lac and is intrinsically of the flat spectrum radio quasar (FSRQ) type, meaning that it possesses an efficient accretion disk and that a dense ultraviolet photon field surrounds the black hole (the Broad Line Region). Such photon fields are optimal for the production of $\gtrsim$~PeV-energy neutrinos. 

Several groups modelled this source, and independently concluded that the maximum neutrino emission during the six-month-long flare must have yielded $N_{\nu_{\mu}+\bar{\nu}_{\mu}} \leq 0.05$ in IceCube~\cite{Ahnen:2018mvi,Cerruti:2018tmc,Gao:2018mnu,Keivani:2018rnh}. The upper limit is a consequence of the EM emission produced in $p\gamma$ interactions, which constitutes approximately 5/8ths of the total energy lost by protons in $p\gamma$ interactions as discussed in Section~\ref{sec:production}. Higher neutrino emission is not compatible with the observed EM emission of the source in the above-quoted models. The upper limit can be somewhat relaxed in the presence of multiple emission zones along the jet~\cite{Xue:2019txw}, or if matter obscures some of the EM emission~\cite{Liu:2018utd}. 

The upper limit of $N_{\nu_{\mu}+\bar{\nu}_{\mu}} \leq 0.05$ means that we expect to observe one out of every 20 equally powerful neutrino flares on average~\cite{Strotjohann:2018ufz}. However, the energetic requirements are extreme in order to achieve $N_{\nu_{\mu}+\bar{\nu}_{\mu}} = 0.05$ in six months. The proton luminosity must be very high, with $\xi > 1000$. Such values of $\xi$ and of $L_p$ are not representative of the blazar population~\cite{Aartsen:2016ngq}. 

For the archival neutrino flare of \txs, most authors do not find a model that can produce sufficient neutrino flux to explain the observations~\cite{Rodrigues:2018tku,Petropoulou:2019zqp}, see also~\cite{Reimer:2018vvw}. But if certain conditions, which include a dense second, slower, jet layer surrounding the faster inner jet are present in \txs, such neutrino flux could be achieved~\cite{Zhang:2019htg}. 

\subsection{\hsp} 

\ichsp arrived from a direction where $90\%$ of neutrinos have energy $0.33^{+2.23}_{-0.27}$ PeV assuming an $E^{-2}$ neutrino spectrum. The extreme blazar \hsp (z = 0.557) is inside the $90\%$ containment region of the neutrino arrival direction. Target-of-opportunity observations revealed high, very hard, and variable X-ray emission from the source shortly after the neutrino arrival~\cite{Giommi:2020viy,Paliya:2021qoa}. The X-ray high state lasted 44 days. 

The leptohadronic modelling of~\cite{Petropoulou:2020pqh} revealed that the Poisson probability to detect one neutrino from the source during the ten years of IceCube's operation is $\sim3\%$ for the most optimistic model studied, while detection of one neutrino during the 44-day-long high X-ray flux-state period following the neutrino detection is lower. The most promising scenarios for neutrino production also predict strong intra-source \gRay attenuation above 100~GeV. If the association is real, then IceCube-Gen2 and other future detectors should be able to provide additional evidence for neutrino production in other extreme blazars with similar \gRay behaviour. 

\subsection{\pks} 

\pks is a very luminous source at large redshift ($z = 1.84$). During 2008-9 it underwent a strong \gRay outburst. During that time, \pks was the second brightest extragalactic \gRay source. \icpks has most probable energy $\sim300$ TeV. At the time of arrival of \icpks, the source was quiet in the \gRay energy range. 

Models of \pks can easily account for the detection of several neutrinos during the lifetime of IceCube, if the emitting region of the neutrinos and \gRays is close to the SMBH which is surrounded by dense radiation fields~\cite{Rodrigues:2020fbu}, 
related to the powerful accretion disk that it possesses. However, radio interferometric and \gRay observations suggest that the emitting region is more likely at a larger distance to the base of the jet~\cite{Karamanavis_b}. Taking these observations into account, we showed in~\cite{Oikonomou2021}, that \pks can have produced up to of order $N_{\nu_{\mu}+\bar{\nu}_{\mu}}\sim0.1$ neutrinos during the ten-year operation of IceCube. An appealing feature of this model is that the required proton luminosity is consistent with the upper limits on the baryon loading factor of the entire blazar population. 

\begin{figure}
\begin{center} 
\includegraphics[trim=0.5cm 0.5cm 0.5cm 0.5cm, clip, width = 0.67\textwidth]{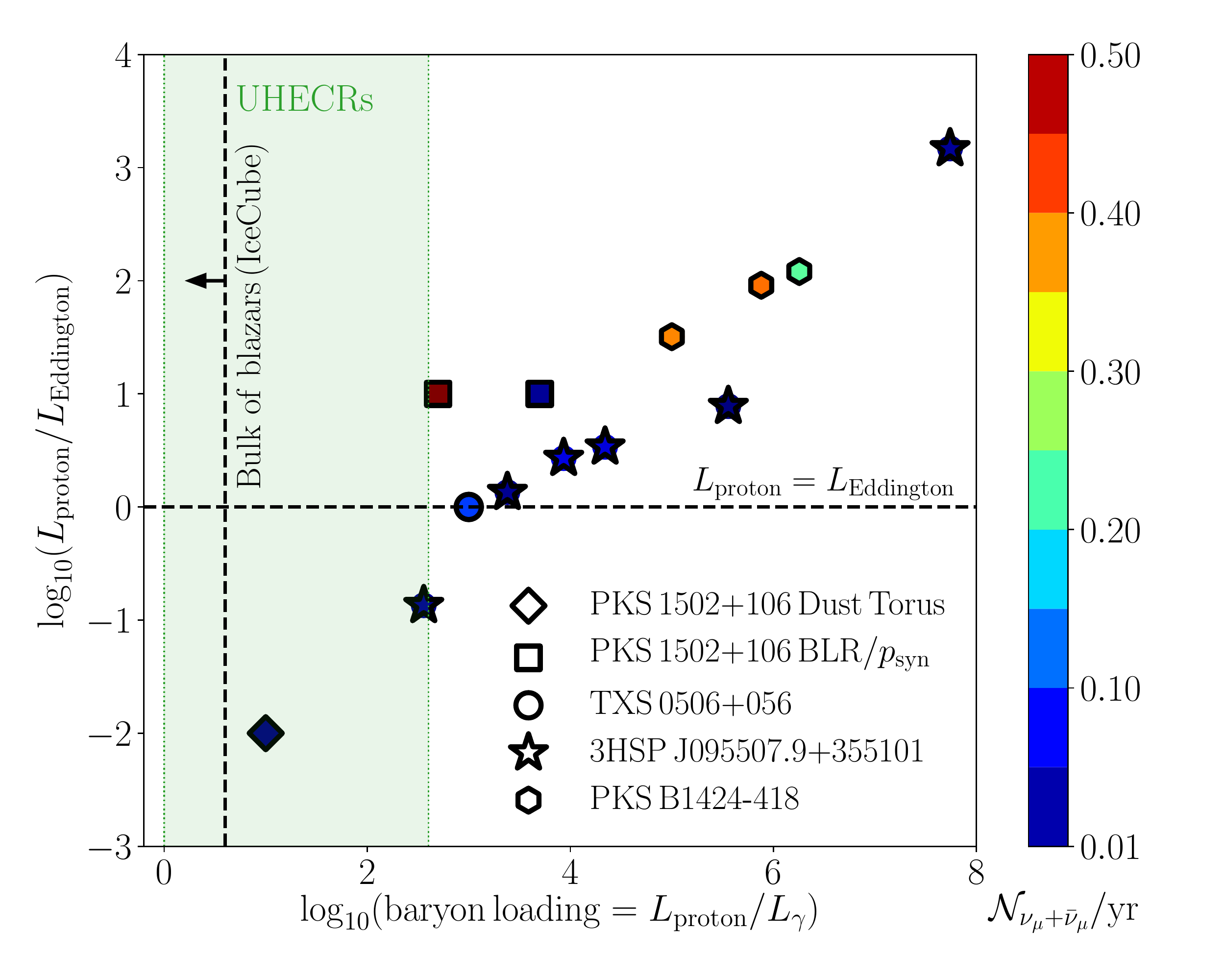}
\vspace{-0.3cm}
\caption{\label{fig:baryonLoad} Baryon loading factor, defined as the ratio of the proton luminosity to the total electromagnetic radiation of the jet, versus ratio of proton to Eddington luminosity in the most optimistic leptohadronic models of \pks, \pksb, \hsp, and \txs. The model predictions listed are from~\cite{Oikonomou2021}~(\pks longterm emission, dust torus model-blue diamond), 
\cite{Rodrigues:2020fbu}~(\pks longterm emission, broad line region model-red square, and proton synchrotron model-blue square), \cite{Keivani:2018rnh}~(\txs longterm emission and emission during 2017 flare - blue circle), \cite{Gao:2016uld}~(\pksb longterm emission and emission during 2014 flare-hexagons), and \cite{Petropoulou:2020pqh}~(\hsp X-ray flare and longterm emission-stars). Models above the horizontal dashed line assume proton luminosity which exceeds the Eddington luminosity of the SMBH. The vertical dashed line denotes the upper limit on the value of the baryon loading factor obtained by the EHE IceCube analysis for the entire blazar population~\cite{2016PhRvL.117x1101A} (recently updated in~\cite{IceCube:2021uhz}) based on the model of~\cite{2014PhRvD..90b3007M}. The green band indicates the values of the baryon loading factor with which jetted AGN reproduce the observed UHECR spectrum~\cite{Rodrigues:2020pli}. }
\vspace{-0.7cm}
\end{center}
\end{figure}

\subsection{Implications} 

\noindent Fig.\,\ref{fig:baryonLoad} summarises the required proton content of the blazar jets in the three aforementioned blazars while maximising the expected number of neutrinos which is in the range 0.01 - 0.5 / year. Additionally, the modelling results of~\cite{Gao:2016uld} on \pksb, which was found to be coincident with the cascade event IC-35 are shown~\cite{Kadler:2016ygj}. The abscissa shows the required baryon-loading factor for a given neutrino flux to have been produced while the ordinate shows the required proton luminosity for each of the sources in terms of the Eddington luminosity of the SMBH. The majority of the models require super-Eddington proton luminosity and baryon loading which is larger than what is inferred from the level of the diffuse neutrino flux measured with IceCube. The baryon-loading factor of extragalactic high-energy sources is independently constrained from UHECR observations. Fits to the UHECR data assuming that they all originate in blazar jets allow $\xi < 400$, which is lower than what is required to account for the neutrino flux implied by the aforementioned blazar-alert-neutrino associations~\cite{2014PhRvD..90b3007M,Rodrigues:2020pli}. On the other hand, the neutrino models summarised in Fig.\,\ref{fig:baryonLoad} above are not directly constrained by UHECR observations, as they don't assume cosmic-ray acceleration to ultra-high energies. 

Such proton luminosity implies, additionally, that the power of the relativistic jet is dominated by the power of relativistic protons and that the total jet power is one to two orders of magnitude higher than was previously inferred for the bulk of blazars~\cite{Ghisellini:2014pwa} as discussed in~\cite{Ghisellini:2019uxc,Petropoulou:2019zqp}. Such enormous jet power can only be accommodated if the accretion efficiency in the presence of a disk with a particular radiative luminosity was previously overestimated in standard accretion models. All in all, if the studied associations that point to super-Eddington proton luminosities are real, they are pointing to extreme events in the lifetime of the blazars. However, the required proton power seems unnaturally high and is in tension with the population-wide limits inferred from IceCube observations. 

\section{Discussion and Outlook} 
\label{sec:discussion} 

\begin{figure}
\begin{center}
\includegraphics[trim=0.2cm 0.0cm 0.5cm 0.5cm, clip, width = \textwidth]{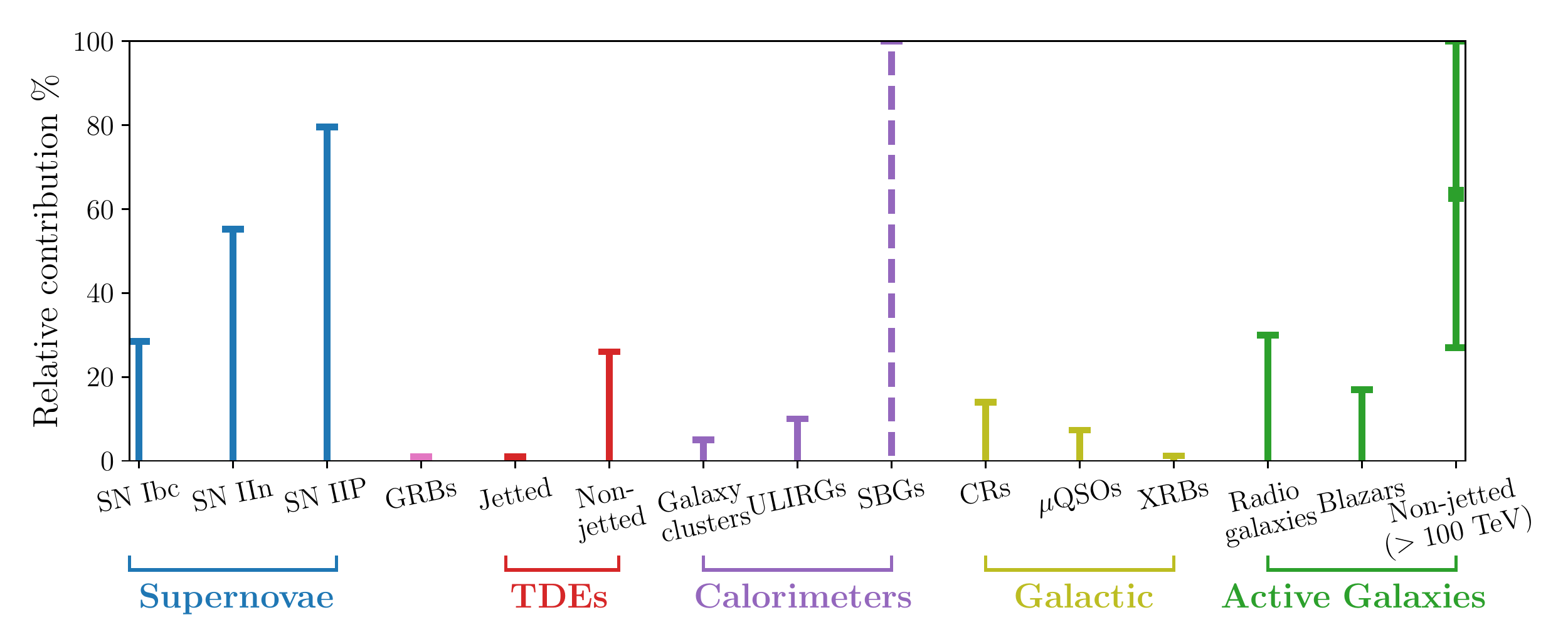}
\vspace{-0.3cm}
\caption{Bar chart of the possible contribution of different astrophysical source classes to the astrophysical neutrino flux measured by IceCube based on stacking analyses. The quoted values are $90\%$ CL upper limits, except in the case of non-jetted AGN where the stacking analysis has led to a $2.6\sigma$ signal and where the flux-fraction implied if the signal is genuine and associated $1\sigma$ uncertainty are shown. See main text for details and for some of the caveats of these results.}
\label{fig:bar_chart} 
\vspace{-0.7cm}
\end{center} 
\end{figure}

\noindent With blazars constrained as the main sources of the high-energy neutrino flux, the quest for the dominant neutrino source class continues. Fig.\,\ref{fig:bar_chart} summarises the possible contribution of different astrophysical source classes to the astrophysical neutrino flux based on the results of stacking analyses. Most of these analyses have resulted in upper limits. The upper limits are obtained based on different assumptions about the relative scaling of the neutrino flux with the EM fluxes of the source populations and are thus, necessarily, heterogeneous. The different analyses 
further suffer from heterogeneous choices of the minimum and maximum neutrino energy and of the assumed spectral index of the neutrino spectrum of the tested sources, which the conclusions of the analyses are very sensitive to. They should thus only be viewed as indicative. For source classes which are numerous with a positive redshift evolution, such as supernovae, catalogue completeness is likely very poor and thus the results of the stacking analyses of those objects should be considered as highly uncertain. 

The figure summarises $90\%$ CL upper limits to the contribution of different source classes to the IceCube neutrino flux. The total contribution of different core-collapse supernovae (SNe) was constrained to be below  $<55\%$ (SNe IIn),  $<80\%$ (SNe IIP) and $<29\%$ (SNe Ibc) of the total diffuse IceCube flux in~\cite{IceCube:2021oiv}, see also~\cite{2018PhDT.......322S}. Core-collapse SNe could still produce $100\%$ of the IceCube flux by summing the above upper limits, and the uncertainties related to catalogue completeness are severe for this source class, see also e.g.~\cite{IceCube:2018omy,Bartos:2021tok}. Therefore, SNe are not constrained as the dominant sources of the IceCube neutrino flux. The prompt phase of \gRay bursts (GRBs) has been constrained to account for $<1\%$ of the high-energy neutrino flux~\cite{IceCube:2016ipa}. Tidal disruption events (TDEs) produce no more than $26\%$ of the high-energy neutrino intensity for non-jetted TDEs and $<1\%$ for jetted-TDEs~\cite{Stein:2019ivm}. Stacking upper limits have been obtained for various calorimetric environments, with ultra-luminous infrared galaxies (ULIRGs) constrained to $<10\%$~\cite{ICECUBE:2021edr}, and galaxy clusters out to redshift, $z=2$ to $<5\%$~\cite{IceCube:2021abh}. The contribution of starburst galaxies (SBGs) has not been constrained by stacking analyses. Being a populous and strongly evolving source population, SBGs are difficult to constrain and by most arguments they are consistent with producing $100\%$ of high-energy neutrinos (see however~\cite{Bartos:2021tok} for a different conclusion). Other stacking analyses constrain the contribution of Galactic sources to the high-energy neutrino flux. Galactic cosmic rays (CRs) can account for no more than $<14\%$~\cite{2017ApJ...849...67A}, microquasars ($\mu$QSOs) $<7\%$, and TeV detected X-ray binaries (XRBs) $<1\%$~\cite{IceCube:2021vlt}. Finally for AGN, stacking analyses constrain the relative contribution of blazars to $<17\%$~\cite{Huber:2019lrm}, and radio galaxies to $<30\%$ of the diffuse neutrino flux~\cite{Zhou:2021rhl}. Recently, a $2.6\sigma$ excess of neutrinos from the directions of non-jetted AGN has been reported~\cite{IceCube:2021pgw}. If this is a genuine signal, it implies that 27\% to 100\% of neutrinos at 100~TeV come from non-jetted AGN. 

Blazars are the first well-studied extragalactic \gRay emitting population and as such the most well-known class of extragalactic high-energy particle accelerators. All in all, the shape of the astrophysical neutrino spectrum, stacking limits, lack of point sources and absence of strong small-scale clustering point against blazars as the dominant source of high-energy astrophysical neutrinos. Theoretically, it is natural to expect that the blazar neutrino spectrum peaks beyond 10~PeV, as summarised in Section~\ref{sec:production}. Therefore, planned neutrino telescopes with peak sensitivity in the sub-EeV energy range such as Ashra-NTA, IceCube-Gen2 Radio, GRAND, POEMMA, RNO-G, Trinity, or Top-of-Mountain fluorescence telescope system will also probe, and possibly characterise the diffuse blazar neutrino flux~\cite{Sasaki:2017uta,IceCube-Gen2:2020qha,GRAND:2018iaj,POEMMA:2020ykm,RNO-G:2020rmc,Brown:2021ane,Neronov:2019htv}. 
 
Blazars remain testable neutrino point sources, especially during flares. Experimentally, short-duration flares are optimal for establishing neutrino emission from a point source. However, for the sources studied in detail thus far, it is not expected to detect sufficient neutrinos during a short-lived flare with a Gton neutrino detector, as the probability to detect neutrinos grows linearly with the neutrino fluence. Therefore, in the coming years, long-lived blazar flares will be most promising for establishing neutrino emission from blazars and thus the acceleration of high-energy hadrons in large-scale relativistic jets. 

\section{Acknowledgements} 
\noindent I would like to thank the ICRC organisers for the nice conference and for the invitation to give this talk. Thanks to Kohta Murase and Michael Unger for constructive comments on this manuscript. I also acknowledge my collaborators for the joint work that was summarised in this contribution. 

\clearpage
\begin{multicols}{2}
\setlength\columnsep{2pt}
\footnotesize
\setlength{\parskip}{-0.3ex}
\setlength{\itemsep}{-1pt}
\bibliography{icrc21}

\hyphenation{Post-Script Sprin-ger}\hyphenation{Post-Script Sprin-ger}

\providecommand{\href}[2]{#2}\begingroup\raggedright\begin{thebibliography}{100}

\bibitem{Aartsen:2013bka}
{\scshape IceCube} collaboration, \emph{{First observation of PeV-energy
  neutrinos with IceCube}},
  \href{https://doi.org/10.1103/PhysRevLett.111.021103}{\emph{Phys.Rev.Lett.}
  {\bfseries 111} (2013) 021103}
  [\href{https://arxiv.org/abs/1304.5356}{{\ttfamily 1304.5356}}].

\bibitem{IceCube:2020acn}
{\scshape IceCube} collaboration, \emph{{Characteristics of the diffuse
  astrophysical electron and tau neutrino flux with six years of IceCube high
  energy cascade data}},
  \href{https://doi.org/10.1103/PhysRevLett.125.121104}{\emph{Phys. Rev. Lett.}
  {\bfseries 125} (2020) 121104}
  [\href{https://arxiv.org/abs/2001.09520}{{\ttfamily 2001.09520}}].

\bibitem{IceCube:2021uhz}
{\scshape IceCube} collaboration, \emph{{Improved Characterization of the
  Astrophysical Muon-Neutrino Flux with 9.5 Years of IceCube Data}},
  \href{https://arxiv.org/abs/2111.10299}{{\ttfamily 2111.10299}}.

\bibitem{IceCube:2018dnn}
{\scshape IceCube, Fermi-LAT, MAGIC, AGILE, ASAS-SN, HAWC, H.E.S.S., INTEGRAL,
  Kanata, Kiso, Kapteyn, Liverpool Telescope, Subaru, Swift NuSTAR, VERITAS,
  VLA/17B-403} collaboration, \emph{{Multimessenger observations of a flaring
  blazar coincident with high-energy neutrino IceCube-170922A}},
  \href{https://doi.org/10.1126/science.aat1378}{\emph{Science} {\bfseries 361}
  (2018) eaat1378} [\href{https://arxiv.org/abs/1807.08816}{{\ttfamily
  1807.08816}}].

\bibitem{Stecker:1978ah}
F.~W. Stecker, \emph{{Diffuse Fluxes of Cosmic High-Energy Neutrinos}},
  \href{https://doi.org/10.1086/156919}{\emph{Astrophys.J.} {\bfseries 228}
  (1979) 919}.

\bibitem{Berezinsky:1992wr}
V.~Berezinsky, T.~Gaisser, F.~Halzen and T.~Stanev, \emph{{Diffuse radiation
  from cosmic ray interactions in the galaxy}},
  \href{https://doi.org/10.1016/0927-6505(93)90014-5}{\emph{Astropart.Phys.}
  {\bfseries 1} (1993) 281}.

\bibitem{PhysRevLett.66.2697}
F.~W. Stecker, C.~Done, M.~H. Salamon and P.~Sommers, \emph{High-energy
  neutrinos from active galactic nuclei},
  \href{https://doi.org/10.1103/PhysRevLett.66.2697}{\emph{Phys. Rev. Lett.}
  {\bfseries 66} (1991) 2697}.

\bibitem{Meszaros:2001ms}
P.~M\'esz\'aros and E.~Waxman, \emph{{TeV neutrinos from successful and choked
  gamma-ray bursts}},
  \href{https://doi.org/10.1103/PhysRevLett.87.171102}{\emph{Phys.Rev.Lett.}
  {\bfseries 87} (2001) 171102}
  [\href{https://arxiv.org/abs/astro-ph/0103275}{{\ttfamily
  astro-ph/0103275}}].

\bibitem{Loeb:2006tw}
A.~Loeb and E.~Waxman, \emph{{The Cumulative background of high energy
  neutrinos from starburst galaxies}},
  \href{https://doi.org/10.1088/1475-7516/2006/05/003}{\emph{JCAP} {\bfseries
  0605} (2006) 003} [\href{https://arxiv.org/abs/astro-ph/0601695}{{\ttfamily
  astro-ph/0601695}}].

\bibitem{Ajello:2013lka}
M.~Ajello, R.~Romani, D.~Gasparrini, M.~Shaw, J.~Bolmer et~al., \emph{{The
  Cosmic Evolution of Fermi BL Lacertae Objects}},
  \href{https://doi.org/10.1088/0004-637X/780/1/73}{\emph{Astrophys.J.}
  {\bfseries 780} (2014) 73} [\href{https://arxiv.org/abs/1310.0006}{{\ttfamily
  1310.0006}}].

\bibitem{1992A&A...260L...1M}
K.~{Mannheim}, T.~{Stanev} and P.~L. {Biermann}, \emph{{Neutrinos from
  flat-spectrum radio quasars}}, {\emph{Astron. Astrophys.} {\bfseries 260}
  (1992) L1}.

\bibitem{1993A&A...269...67M}
K.~{Mannheim}, \emph{{The proton blazar.}}, {\emph{Astron. Astrophys.}
  {\bfseries 269} (1993) 67}
  [\href{https://arxiv.org/abs/astro-ph/9302006}{{\ttfamily
  astro-ph/9302006}}].

\bibitem{Atoyan:2001ey}
A.~Atoyan and C.~D. Dermer, \emph{{High-energy neutrinos from photomeson
  processes in blazars}},
  \href{https://doi.org/10.1103/PhysRevLett.87.221102}{\emph{Phys.Rev.Lett.}
  {\bfseries 87} (2001) 221102}
  [\href{https://arxiv.org/abs/astro-ph/0108053}{{\ttfamily
  astro-ph/0108053}}].

\bibitem{Muecke:2002bi}
A.~Muecke, R.~J. Protheroe, R.~Engel, J.~P. Rachen and T.~Stanev, \emph{{BL Lac
  Objects in the synchrotron proton blazar model}},
  \href{https://doi.org/10.1016/S0927-6505(02)00185-8}{\emph{Astropart. Phys.}
  {\bfseries 18} (2003) 593}
  [\href{https://arxiv.org/abs/astro-ph/0206164}{{\ttfamily
  astro-ph/0206164}}].

\bibitem{2003ApJ...586...79A}
A.~M. {Atoyan} and C.~D. {Dermer}, \emph{{Neutral Beams from Blazar Jets}},
  \href{https://doi.org/10.1086/346261}{\emph{Astrophys. J.} {\bfseries 586}
  (2003) 79} [\href{https://arxiv.org/abs/astro-ph/0209231}{{\ttfamily
  astro-ph/0209231}}].

\bibitem{Neronov:2002xv}
A.~Y. Neronov and D.~V. Semikoz, \emph{{Which blazars are neutrino loud?}},
  \href{https://doi.org/10.1103/PhysRevD.66.123003}{\emph{Phys. Rev. D}
  {\bfseries 66} (2002) 123003}
  [\href{https://arxiv.org/abs/hep-ph/0208248}{{\ttfamily hep-ph/0208248}}].

\bibitem{Dermer:2007me}
C.~Dermer, E.~Ramirez-Ruiz and T.~Le, \emph{{Correlation of Photon and Neutrino
  Fluxes in Blazars and Gamma Ray Bursts}},
  \href{https://doi.org/10.1086/520638}{\emph{Astrophys.J.} {\bfseries 664}
  (2007) L67} [\href{https://arxiv.org/abs/astro-ph/0703219}{{\ttfamily
  astro-ph/0703219}}].

\bibitem{2014PhRvD..90b3007M}
K.~{Murase}, Y.~{Inoue} and C.~D. {Dermer}, \emph{{Diffuse neutrino intensity
  from the inner jets of active galactic nuclei: Impacts of external photon
  fields and the blazar sequence}},
  \href{https://doi.org/10.1103/PhysRevD.90.023007}{\emph{Phys. Rev. D}
  {\bfseries 90} (2014) 023007}
  [\href{https://arxiv.org/abs/1403.4089}{{\ttfamily 1403.4089}}].

\bibitem{Tavecchio:2014xha}
F.~Tavecchio, G.~Ghisellini and D.~Guetta, \emph{{Structured Jets in BL Lac
  Objects: Efficient PeV Neutrino Factories?}},
  \href{https://doi.org/10.1088/2041-8205/793/1/L18}{\emph{Astrophys.J.}
  {\bfseries 793} (2014) L18}.

\bibitem{Petropoulou:2015upa}
M.~Petropoulou, S.~Dimitrakoudis, P.~Padovani, A.~Mastichiadis and E.~Resconi,
  \emph{{Photohadronic origin of $\boldsymbol {\gamma }$-ray BL Lac emission:
  implications for IceCube neutrinos}},
  \href{https://doi.org/10.1093/mnras/stv179}{\emph{MNRAS} {\bfseries 448}
  (2015) 2412} [\href{https://arxiv.org/abs/1501.07115}{{\ttfamily
  1501.07115}}].

\bibitem{Petropoulou:2016ujj}
M.~Petropoulou, S.~Coenders and S.~Dimitrakoudis, \emph{{Time-dependent
  neutrino emission from Mrk 421 during flares and predictions for IceCube}},
  \href{https://doi.org/10.1016/j.astropartphys.2016.04.001}{\emph{Astropart.
  Phys.} {\bfseries 80} (2016) 115}
  [\href{https://arxiv.org/abs/1603.06954}{{\ttfamily 1603.06954}}].

\bibitem{Padovani:2015mba}
P.~Padovani, M.~Petropoulou, P.~Giommi and E.~Resconi, \emph{{A simplified view
  of blazars: the neutrino background}},
  \href{https://doi.org/10.1093/mnras/stv1467}{\emph{MNRAS} {\bfseries 452}
  (2015) 1877} [\href{https://arxiv.org/abs/1506.09135}{{\ttfamily
  1506.09135}}].

\bibitem{Gao:2016uld}
S.~Gao, M.~Pohl and W.~Winter, \emph{{On the direct correlation between
  gamma-rays and PeV neutrinos from blazars}},
  \href{https://doi.org/10.3847/1538-4357/aa7754}{\emph{Astrophys. J.}
  {\bfseries 843} (2017) 109}
  [\href{https://arxiv.org/abs/1610.05306}{{\ttfamily 1610.05306}}].

\bibitem{Rodrigues:2017fmu}
X.~Rodrigues, A.~Fedynitch, S.~Gao, D.~Boncioli and W.~Winter, \emph{{Neutrinos
  and Ultra-High-Energy Cosmic-Ray Nuclei from Blazars}},
  \href{https://doi.org/10.3847/1538-4357/aaa7ee}{\emph{Astrophys. J.}
  {\bfseries 854} (2018) 54}
  [\href{https://arxiv.org/abs/1711.02091}{{\ttfamily 1711.02091}}].

\bibitem{Palladino:2018lov}
A.~Palladino, X.~Rodrigues, S.~Gao and W.~Winter, \emph{{Interpretation of the
  diffuse astrophysical neutrino flux in terms of the blazar sequence}},
  \href{https://doi.org/10.3847/1538-4357/aaf507}{\emph{Astrophys. J.}
  {\bfseries 871} (2019) 41}
  [\href{https://arxiv.org/abs/1806.04769}{{\ttfamily 1806.04769}}].

\bibitem{Rodrigues:2020pli}
X.~Rodrigues, J.~Heinze, A.~Palladino, A.~van Vliet and W.~Winter,
  \emph{{Active Galactic Nuclei Jets as the Origin of Ultrahigh-Energy Cosmic
  Rays and Perspectives for the Detection of Astrophysical Source Neutrinos at
  EeV Energies}},
  \href{https://doi.org/10.1103/PhysRevLett.126.191101}{\emph{{Phys. Rev.
  Lett.}} {\bfseries 126} (2021) 191101}
  [\href{https://arxiv.org/abs/2003.08392}{{\ttfamily 2003.08392}}].

\bibitem{Murase:2018iyl}
K.~Murase, F.~Oikonomou and M.~Petropoulou, \emph{{Blazar Flares as an Origin
  of High-Energy Cosmic Neutrinos?}},
  \href{https://doi.org/10.3847/1538-4357/aada00}{\emph{Astrophys. J.}
  {\bfseries 865} (2018) 124}
  [\href{https://arxiv.org/abs/1807.04748}{{\ttfamily 1807.04748}}].

\bibitem{Waxman:1998yy}
E.~Waxman and J.~N. Bahcall, \emph{{High-energy neutrinos from astrophysical
  sources: An Upper bound}},
  \href{https://doi.org/10.1103/PhysRevD.59.023002}{\emph{Phys.Rev.} {\bfseries
  D59} (1998) 023002} [\href{https://arxiv.org/abs/hep-ph/9807282}{{\ttfamily
  hep-ph/9807282}}].

\bibitem{Murase:2015xka}
K.~Murase, D.~Guetta and M.~Ahlers, \emph{{Hidden Cosmic-Ray Accelerators as an
  Origin of TeV-PeV Cosmic Neutrinos}},
  \href{https://doi.org/10.1103/PhysRevLett.116.071101}{\emph{{Phys. Rev.
  Lett.}} {\bfseries 116} (2016) 071101}
  [\href{https://arxiv.org/abs/1509.00805}{{\ttfamily 1509.00805}}].

\bibitem{petro_2015}
M.~{Petropoulou} and A.~{Mastichiadis}, \emph{{Bethe-Heitler emission in BL
  Lacs: filling the gap between X-rays and {$\gamma$}-rays}},
  \href{https://doi.org/10.1093/mnras/stu2364}{\emph{MNRAS} {\bfseries 447}
  (2015) 36} [\href{https://arxiv.org/abs/1411.1908}{{\ttfamily 1411.1908}}].

\bibitem{Ahnen:2018mvi}
{\scshape MAGIC} collaboration, \emph{{The blazar TXS 0506+056 associated with
  a high-energy neutrino: insights into extragalactic jets and cosmic ray
  acceleration}},
  \href{https://doi.org/10.3847/2041-8213/aad083}{\emph{Astrophys. J. Lett.}
  {\bfseries 863} (2018) L10}
  [\href{https://arxiv.org/abs/1807.04300}{{\ttfamily 1807.04300}}].

\bibitem{Cerruti:2018tmc}
M.~Cerruti, A.~Zech, C.~Boisson, G.~Emery, S.~Inoue and J.~P. Lenain,
  \emph{{Lepto-hadronic single-zone models for the electromagnetic and neutrino
  emission of TXS 0506+056}},
  \href{https://doi.org/10.1093/mnrasl/sly210}{\emph{MNRAS} {\bfseries 483}
  (2019) L12} [\href{https://arxiv.org/abs/1807.04335}{{\ttfamily
  1807.04335}}].

\bibitem{Gao:2018mnu}
S.~Gao, A.~Fedynitch, W.~Winter and M.~Pohl, \emph{{Modelling the coincident
  observation of a high-energy neutrino and a bright blazar flare}},
  \href{https://doi.org/10.1038/s41550-018-0610-1}{\emph{Nat. Astron.}
  {\bfseries 3} (2019) 88} [\href{https://arxiv.org/abs/1807.04275}{{\ttfamily
  1807.04275}}].

\bibitem{Keivani:2018rnh}
A.~Keivani et~al., \emph{{A Multimessenger Picture of the Flaring Blazar TXS
  0506+056: implications for High-Energy Neutrino Emission and Cosmic Ray
  Acceleration}},
  \href{https://doi.org/10.3847/1538-4357/aad59a}{\emph{Astrophys. J.}
  {\bfseries 864} (2018) 84}
  [\href{https://arxiv.org/abs/1807.04537}{{\ttfamily 1807.04537}}].

\bibitem{2019ICRC...36.1017S}
{\scshape IceCube} collaboration, \emph{{Measurement of the Diffuse
  Astrophysical Muon-Neutrino Spectrum with Ten Years of IceCube Data}},
  \href{https://doi.org/10.22323/1.358.1017}{\emph{PoS} {\bfseries ICRC2019}
  (2020) 1017} [\href{https://arxiv.org/abs/1908.09551}{{\ttfamily
  1908.09551}}].

\bibitem{Aartsen:2018vtx}
{\scshape IceCube} collaboration, \emph{{Differential limit on the
  extremely-high-energy cosmic neutrino flux in the presence of astrophysical
  background from nine years of IceCube data}},
  \href{https://doi.org/10.1103/PhysRevD.98.062003}{\emph{Phys. Rev.}
  {\bfseries D98} (2018) 062003}
  [\href{https://arxiv.org/abs/1807.01820}{{\ttfamily 1807.01820}}].

\bibitem{Zas:2017Yb}
E.~Zas, \emph{{Searches for neutrino fluxes in the EeV regime with the Pierre
  Auger Observatory}}, \href{https://doi.org/10.22323/1.301.0972}{\emph{PoS}
  {\bfseries ICRC2017} (2017) 972}.

\bibitem{Huber:2019lrm}
{\scshape IceCube} collaboration, \emph{{Searches for steady neutrino emission
  from 3FHLblazars using eight years of IceCube data from theNorthern
  hemisphere}}, \href{https://doi.org/10.22323/1.358.0916}{\emph{PoS}
  {\bfseries ICRC2019} (2020) 916}
  [\href{https://arxiv.org/abs/1908.08458}{{\ttfamily 1908.08458}}].

\bibitem{IceCube:2020wum}
{\scshape IceCube} collaboration, \emph{{The IceCube high-energy starting event
  sample: Description and flux characterization with 7.5 years of data}},
  \href{https://doi.org/10.1103/PhysRevD.104.022002}{\emph{Phys. Rev. D}
  {\bfseries 104} (2021) 022002}
  [\href{https://arxiv.org/abs/2011.03545}{{\ttfamily 2011.03545}}].

\bibitem{TheFermi-LAT:2017pvy}
{\scshape Fermi-LAT} collaboration, \emph{{3FHL: The Third Catalog of Hard
  Fermi-LAT Sources}},
  \href{https://doi.org/10.3847/1538-4365/aa8221}{\emph{Astrophys. J., Suppl.
  Ser.} {\bfseries 232} (2017) 18}
  [\href{https://arxiv.org/abs/1702.00664}{{\ttfamily 1702.00664}}].

\bibitem{Murase:2016gly}
K.~Murase and E.~Waxman, \emph{{Constraining High-Energy Cosmic Neutrino
  Sources: Implications and Prospects}},
  \href{https://doi.org/10.1103/PhysRevD.94.103006}{\emph{Phys. Rev.}
  {\bfseries D94} (2016) 103006}
  [\href{https://arxiv.org/abs/1607.01601}{{\ttfamily 1607.01601}}].

\bibitem{Ando:2017xcb}
S.~Ando, M.~R. Feyereisen and M.~Fornasa, \emph{{How bright can the brightest
  neutrino source be?}},
  \href{https://doi.org/10.1103/PhysRevD.95.103003}{\emph{Phys. Rev.}
  {\bfseries D95} (2017) 103003}
  [\href{https://arxiv.org/abs/1701.02165}{{\ttfamily 1701.02165}}].

\bibitem{Yuan:2019ucv}
C.~{Yuan}, K.~{Murase} and P.~{M{\'e}sz{\'a}ros}, \emph{{Complementarity of
  Stacking and Multiplet Constraints on the Blazar Contribution to the
  Cumulative High-energy Neutrino Intensity}},
  \href{https://doi.org/10.3847/1538-4357/ab65ea}{\emph{Astrophys. J.}
  {\bfseries 890} (2020) 25}
  [\href{https://arxiv.org/abs/1904.06371}{{\ttfamily 1904.06371}}].

\bibitem{Capel:2020txc}
F.~Capel, D.~J. Mortlock and C.~Finley, \emph{{Bayesian constraints on the
  astrophysical neutrino source population from IceCube data}},
  \href{https://doi.org/10.1103/PhysRevD.101.123017}{\emph{Phys. Rev. D}
  {\bfseries 101} (2020) 123017}
  [\href{https://arxiv.org/abs/2005.02395}{{\ttfamily 2005.02395}}].

\bibitem{Waxman:2013zda}
E.~Waxman, \emph{{IceCube's Neutrinos: The beginning of extra-Galactic neutrino
  astrophysics?}},  \href{https://arxiv.org/abs/1312.0558}{{\ttfamily
  1312.0558}}.

\bibitem{Ahlers:2018fkn}
M.~Ahlers and F.~Halzen, \emph{{Opening a New Window onto the Universe with
  IceCube}}, \href{https://doi.org/10.1016/j.ppnp.2018.05.001}{\emph{Prog.
  Part. Nucl. Phys.} {\bfseries 102} (2018) 73}
  [\href{https://arxiv.org/abs/1805.11112}{{\ttfamily 1805.11112}}].

\bibitem{Murase:2018utn}
K.~Murase and M.~Fukugita, \emph{{Energetics of High-Energy Cosmic
  Radiations}}, \href{https://doi.org/10.1103/PhysRevD.99.063012}{\emph{Phys.
  Rev. D} {\bfseries 99} (2019) 063012}
  [\href{https://arxiv.org/abs/1806.04194}{{\ttfamily 1806.04194}}].

\bibitem{TheFermi-LAT:2015ykq}
{\scshape Fermi LAT Collaboration} collaboration, \emph{{Resolving the
  Extragalactic $\gamma$-Ray Background above 50 GeV with the Fermi Large Area
  Telescope}},
  \href{https://doi.org/10.1103/PhysRevLett.116.151105}{\emph{{Phys. Rev.
  Lett.}} {\bfseries 116} (2016) 151105}
  [\href{https://arxiv.org/abs/1511.00693}{{\ttfamily 1511.00693}}].

\bibitem{Lisanti:2016jub}
M.~Lisanti, S.~Mishra-Sharma, L.~Necib and B.~R. Safdi, \emph{{Deciphering
  Contributions to the Extragalactic Gamma-Ray Background from 2 GeV to 2
  TeV}}, \href{https://doi.org/10.3847/0004-637X/832/2/117}{\emph{Astrophys.
  J.} {\bfseries 832} (2016) 117}
  [\href{https://arxiv.org/abs/1606.04101}{{\ttfamily 1606.04101}}].

\bibitem{Zechlin:2016pme}
H.-S. Zechlin, A.~Cuoco, F.~Donato, N.~Fornengo and M.~Regis,
  \emph{{Statistical Measurement of the Gamma-ray Source-count Distribution as
  a Function of Energy}},
  \href{https://doi.org/10.3847/2041-8205/826/2/L31}{\emph{Astrophys. J. Lett.}
  {\bfseries 826} (2016) L31}
  [\href{https://arxiv.org/abs/1605.04256}{{\ttfamily 1605.04256}}].

\bibitem{Oikonomou:2019djc}
F.~Oikonomou, K.~Murase, P.~Padovani, E.~Resconi and P.~M\'{e}sz\'{a}ros,
  \emph{{High energy neutrino flux from individual blazar flares}},
  \href{https://doi.org/10.1093/mnras/stz2246}{\emph{MNRAS} {\bfseries 489}
  (2019) 4347} [\href{https://arxiv.org/abs/1906.05302}{{\ttfamily
  1906.05302}}].

\bibitem{Oikonomou2021}
F.~{Oikonomou}, M.~{Petropoulou}, K.~{Murase}, A.~{Tohuvavohu},
  G.~{Vasilopoulos}, S.~{Buson} et~al., \emph{{Multi-messenger emission from
  the parsec-scale jet of the flat-spectrum radio quasar coincident with
  high-energy neutrino IceCube-190730A}},
  \href{https://doi.org/10.1088/1475-7516/2021/10/082}{\emph{JCAP} {\bfseries
  10} (2021) 082} [\href{https://arxiv.org/abs/2107.11437}{{\ttfamily
  2107.11437}}].

\bibitem{Petropoulou:2020pqh}
M.~Petropoulou, F.~Oikonomou, A.~Mastichiadis, K.~Murase, P.~Padovani,
  G.~Vasilopoulos et~al., \emph{{Comprehensive Multimessenger Modeling of the
  Extreme Blazar 3HSP J095507.9+355101 and Predictions for IceCube}},
  \href{https://doi.org/10.3847/1538-4357/aba8a0}{\emph{Astrophys. J.}
  {\bfseries 899} (2020) 113}
  [\href{https://arxiv.org/abs/2005.07218}{{\ttfamily 2005.07218}}].

\bibitem{e-ASTROGAM:2016bph}
{\scshape e-ASTROGAM} collaboration, \emph{{The e-ASTROGAM mission}},
  \href{https://doi.org/10.1007/s10686-017-9533-6}{\emph{Exper. Astron.}
  {\bfseries 44} (2017) 25} [\href{https://arxiv.org/abs/1611.02232}{{\ttfamily
  1611.02232}}].

\bibitem{AMEGO:2019gny}
{\scshape AMEGO} collaboration, \emph{{All-sky Medium Energy Gamma-ray
  Observatory: Exploring the Extreme Multimessenger Universe}},
  \href{https://arxiv.org/abs/1907.07558}{{\ttfamily 1907.07558}}.

\bibitem{cta}
``{CTAO's expected performance}.''
  \url{https://www.cta-observatory.org/science/ctao-performance/}.

\bibitem{Plavin:2020emb}
A.~Plavin, Y.~Y. Kovalev, Y.~A. Kovalev and S.~Troitsky, \emph{{Observational
  Evidence for the Origin of High-energy Neutrinos in Parsec-scale Nuclei of
  Radio-bright Active Galaxies}},
  \href{https://doi.org/10.3847/1538-4357/ab86bd}{\emph{Astrophys. J.}
  {\bfseries 894} (2020) 101}
  [\href{https://arxiv.org/abs/2001.00930}{{\ttfamily 2001.00930}}].

\bibitem{Giommi:2020hbx}
P.~Giommi, T.~Glauch, P.~Padovani, E.~Resconi, A.~Turcati and Y.~L. Chang,
  \emph{{Dissecting the regions around IceCube high-energy neutrinos: growing
  evidence for the blazar connection}},
  \href{https://doi.org/10.1093/mnras/staa2082}{\emph{Mon. Not. Roy. Astron.
  Soc.} {\bfseries 497} (2020) 865}
  [\href{https://arxiv.org/abs/2001.09355}{{\ttfamily 2001.09355}}].

\bibitem{Hovatta:2020lor}
T.~{Hovatta}, E.~{Lindfors}, S.~{Kiehlmann}, W.~{Max-Moerbeck}, M.~{Hodges},
  I.~{Liodakis} et~al., \emph{{Association of IceCube neutrinos with radio
  sources observed at Owens Valley and Mets{\"a}hovi Radio Observatories}},
  \href{https://doi.org/10.1051/0004-6361/202039481}{\emph{Astron. Astrophys.}
  {\bfseries 650} (2021) A83}
  [\href{https://arxiv.org/abs/2009.10523}{{\ttfamily 2009.10523}}].

\bibitem{Franckowiak:2020qrq}
A.~Franckowiak et~al., \emph{{Patterns in the Multiwavelength Behavior of
  Candidate Neutrino Blazars}},
  \href{https://doi.org/10.3847/1538-4357/ab8307}{\emph{Astrophys. J.}
  {\bfseries 893} (2020) 162}
  [\href{https://arxiv.org/abs/2001.10232}{{\ttfamily 2001.10232}}].

\bibitem{IceCube:2018cha}
{\scshape IceCube} collaboration, \emph{{Neutrino emission from the direction
  of the blazar TXS 0506+056 prior to the IceCube-170922A alert}},
  \href{https://doi.org/10.1126/science.aat2890}{\emph{Science} {\bfseries 361}
  (2018) 147} [\href{https://arxiv.org/abs/1807.08794}{{\ttfamily
  1807.08794}}].

\bibitem{Stathopoulos:2021mli}
S.~I. Stathopoulos, M.~Petropoulou, P.~Giommi, G.~Vasilopoulos, P.~Padovani and
  A.~Mastichiadis, \emph{{High-energy neutrinos from X-rays flares of blazars
  frequently observed by the Swift X-Ray Telescope}},
  \href{https://doi.org/10.1093/mnras/stab3404}{\emph{MNRAS} (2021) }
  [\href{https://arxiv.org/abs/2111.09320}{{\ttfamily 2111.09320}}].

\bibitem{Garrappa:2021ihz}
{\scshape Fermi-LAT} collaboration, \emph{{Fermi-LAT realtime follow-ups of
  high-energy neutrino alerts}},
  \href{https://doi.org/10.22323/1.395.0956}{\emph{PoS} {\bfseries ICRC2021}
  (2021) 956} [\href{https://arxiv.org/abs/2112.11586}{{\ttfamily
  2112.11586}}].

\bibitem{Acciari:2021YA}
V.~A. Acciari, S.~Ansoldi, L.~A. Antonelli, A.~Arbet~Engels, M.~Artero,
  K.~Asano et~al., \emph{{Searching for VHE gamma-ray emission associated with
  IceCube neutrino alerts using FACT, H.E.S.S., MAGIC, and VERITAS}},
  \href{https://doi.org/10.22323/1.395.0960}{\emph{PoS} {\bfseries ICRC2021}
  (2021) 960}.

\bibitem{Abbasi:2021XE}
R.~Abbasi, M.~Ackermann, J.~Adams, J.~Aguilar, M.~Ahlers, M.~Ahrens et~al.,
  \emph{{A model-independent analysis of neutrino flares detected in IceCube
  from X-ray selected blazars}},
  \href{https://doi.org/10.22323/1.395.0971}{\emph{PoS} {\bfseries ICRC2021}
  (2021) 971}.

\bibitem{Aublin:2021q6}
J.~Aublin and A.~Plavin, \emph{{Search for an association between neutrinos and
  radio-selected blazars with ANTARES}},
  \href{https://doi.org/10.22323/1.395.1164}{\emph{PoS} {\bfseries ICRC2021}
  (2021) 1164}.

\bibitem{Illuminati:202198}
G.~Illuminati and A.~Plavin, \emph{{ANTARES search for neutrino flares from the
  direction of radio-bright blazars}},
  \href{https://doi.org/10.22323/1.395.0972}{\emph{PoS} {\bfseries ICRC2021}
  (2021) 972}.

\bibitem{Kadler:20212/}
M.~Kadler, U.~Bach, D.~Berge, S.~Buson, D.~Dorner, P.~G. Edwards et~al.,
  \emph{{TELAMON: Monitoring of AGN with the Effelsberg 100-m Telescope in the
  Context of Astroparticle Physics}},
  \href{https://doi.org/10.22323/1.395.0974}{\emph{PoS} {\bfseries ICRC2021}
  (2021) 974}.

\bibitem{Padovani:2019xcv}
P.~Padovani, F.~Oikonomou, M.~Petropoulou, P.~Giommi and E.~Resconi, \emph{{TXS
  0506+056, the first cosmic neutrino source, is not a BL Lac}},
  \href{https://doi.org/10.1093/mnrasl/slz011}{\emph{MNRAS} {\bfseries 484}
  (2019) L104} [\href{https://arxiv.org/abs/1901.06998}{{\ttfamily
  1901.06998}}].

\bibitem{Xue:2019txw}
R.~Xue, R.-Y. Liu, M.~Petropoulou, F.~Oikonomou, Z.-R. Wang, K.~Wang et~al.,
  \emph{{A two-zone model for blazar emission: implications for TXS 0506+056
  and the neutrino event IceCube-170922A}},
  \href{https://doi.org/10.3847/1538-4357/ab4b44}{\emph{Astrophys. J.}
  {\bfseries 886} (2019) 23}
  [\href{https://arxiv.org/abs/1908.10190}{{\ttfamily 1908.10190}}].

\bibitem{Liu:2018utd}
R.-Y. Liu, K.~Wang, R.~Xue, A.~M. Taylor, X.-Y. Wang, Z.~Li et~al.,
  \emph{{Hadronuclear interpretation of a high-energy neutrino event coincident
  with a blazar flare}},
  \href{https://doi.org/10.1103/PhysRevD.99.063008}{\emph{Phys. Rev.}
  {\bfseries D99} (2019) 063008}
  [\href{https://arxiv.org/abs/1807.05113}{{\ttfamily 1807.05113}}].

\bibitem{Strotjohann:2018ufz}
N.~L. Strotjohann, M.~Kowalski and A.~Franckowiak, \emph{{Eddington bias for
  cosmic neutrino sources}},
  \href{https://doi.org/10.1051/0004-6361/201834750}{\emph{{Astron.
  Astrophys.}} {\bfseries 622} (2019) L9}
  [\href{https://arxiv.org/abs/1809.06865}{{\ttfamily 1809.06865}}].

\bibitem{Aartsen:2016ngq}
{\scshape IceCube} collaboration, \emph{{Constraints on Ultrahigh-Energy
  Cosmic-Ray Sources from a Search for Neutrinos above 10 PeV with IceCube}},
  \href{https://doi.org/10.1103/PhysRevLett.117.241101,
  10.1103/PhysRevLett.119.259902}{\emph{{Phys. Rev. Lett.}} {\bfseries 117}
  (2016) 241101} [\href{https://arxiv.org/abs/1607.05886}{{\ttfamily
  1607.05886}}].

\bibitem{Rodrigues:2018tku}
X.~Rodrigues, S.~Gao, A.~Fedynitch, A.~Palladino and W.~Winter,
  \emph{{Leptohadronic blazar models applied to the 2014-15 flare of TXS
  0506+056}}, \href{https://doi.org/10.3847/2041-8213/ab1267}{\emph{Astrophys.
  J.} {\bfseries 874} (2019) L29}
  [\href{https://arxiv.org/abs/1812.05939}{{\ttfamily 1812.05939}}].

\bibitem{Petropoulou:2019zqp}
M.~Petropoulou et~al., \emph{{Multi-Epoch Modeling of TXS 0506+056 and
  Implications for Long-Term High-Energy Neutrino Emission}},
  \href{https://doi.org/10.3847/1538-4357/ab76d0}{\emph{Astrophys. J.}
  {\bfseries 891} (2020) 115}
  [\href{https://arxiv.org/abs/1911.04010}{{\ttfamily 1911.04010}}].

\bibitem{Reimer:2018vvw}
A.~Reimer, M.~Boettcher and S.~Buson, \emph{{Cascading Constraints from
  Neutrino-emitting Blazars: The Case of TXS 0506+056}},
  \href{https://doi.org/10.3847/1538-4357/ab2bff}{\emph{Astrophys. J.}
  {\bfseries 881} (2019) 46}
  [\href{https://arxiv.org/abs/1812.05654}{{\ttfamily 1812.05654}}].

\bibitem{Zhang:2019htg}
B.~T. Zhang, M.~Petropoulou, K.~Murase and F.~Oikonomou, \emph{{A Neutral Beam
  Model for High-Energy Neutrino Emission from the Blazar TXS 0506+056}},
  \href{https://doi.org/10.3847/1538-4357/ab659a}{\emph{Astrophys. J.}
  {\bfseries 889} (2020) 118}
  [\href{https://arxiv.org/abs/1910.11464}{{\ttfamily 1910.11464}}].

\bibitem{Giommi:2020viy}
P.~Giommi, P.~Padovani, F.~Oikonomou, T.~Glauch, S.~Paiano and E.~Resconi,
  \emph{{3HSP J095507.9+355101: a flaring extreme blazar coincident in space
  and time with IceCube-200107A}},
  \href{https://doi.org/10.1051/0004-6361/202038423}{\emph{Astron. Astrophys.}
  {\bfseries 640} (2020) L4}
  [\href{https://arxiv.org/abs/2003.06405}{{\ttfamily 2003.06405}}].

\bibitem{Paliya:2021qoa}
V.~S. Paliya, M.~Boettcher, M.~Gurwell and C.~S. Stalin, \emph{{On the Origin
  of Gamma-Ray Flares from Bright Fermi Blazars}},
  \href{https://doi.org/10.3847/1538-4365/ac365d}{\emph{Astrophys. J. Supp.}
  {\bfseries 257} (2021) 37}
  [\href{https://arxiv.org/abs/2111.04379}{{\ttfamily 2111.04379}}].

\bibitem{Rodrigues:2020fbu}
X.~{Rodrigues}, S.~{Garrappa}, S.~{Gao}, V.~S. {Paliya}, A.~{Franckowiak} and
  W.~{Winter}, \emph{{Multiwavelength and Neutrino Emission from Blazar PKS
  1502 + 106}},
  \href{https://doi.org/10.3847/1538-4357/abe87b}{\emph{Astrophys. J.}
  {\bfseries 912} (2021) 54}
  [\href{https://arxiv.org/abs/2009.04026}{{\ttfamily 2009.04026}}].

\bibitem{Karamanavis_b}
V.~{Karamanavis}, L.~{Fuhrmann}, E.~{Angelakis}, I.~{Nestoras}, I.~{Myserlis},
  T.~P. {Krichbaum} et~al., \emph{{What can the 2008/10 broadband flare of PKS
  1502+106 tell us?. Nuclear opacity, magnetic fields, and the location of
  {\ensuremath{\gamma}} rays}},
  \href{https://doi.org/10.1051/0004-6361/201527796}{\emph{Astron. Astrophys.}
  {\bfseries 590} (2016) A48}
  [\href{https://arxiv.org/abs/1603.04220}{{\ttfamily 1603.04220}}].

\bibitem{2016PhRvL.117x1101A}
M.~G. {Aartsen}, K.~{Abraham}, M.~{Ackermann}, J.~{Adams}, J.~A. {Aguilar},
  M.~{Ahlers} et~al., \emph{{Constraints on Ultrahigh-Energy Cosmic-Ray Sources
  from a Search for Neutrinos above 10 PeV with IceCube}},
  \href{https://doi.org/10.1103/PhysRevLett.117.241101}{\emph{Phys. Rev. Lett.}
  {\bfseries 117} (2016) 241101}
  [\href{https://arxiv.org/abs/1607.05886}{{\ttfamily 1607.05886}}].

\bibitem{Kadler:2016ygj}
M.~Kadler et~al., \emph{{Coincidence of a high-fluence blazar outburst with a
  PeV-energy neutrino event}},
  \href{https://doi.org/10.1038/NPHYS3715}{\emph{Nature Phys.} {\bfseries 12}
  (2016) 807} [\href{https://arxiv.org/abs/1602.02012}{{\ttfamily
  1602.02012}}].

\bibitem{Ghisellini:2014pwa}
G.~Ghisellini, F.~Tavecchio, L.~Maraschi, A.~Celotti and T.~Sbarrato,
  \emph{{The power of relativistic jets is larger than the luminosity of their
  accretion disks}}, \href{https://doi.org/10.1038/nature13856}{\emph{Nature}
  {\bfseries 515} (2014) 376}
  [\href{https://arxiv.org/abs/1411.5368}{{\ttfamily 1411.5368}}].

\bibitem{Ghisellini:2019uxc}
G.~Ghisellini, \emph{{Extra-galactic jets: a hard X-ray view}}, {\emph{Mem.
  Soc. Ast. It.} {\bfseries 90} (2019) 154}
  [\href{https://arxiv.org/abs/1911.11777}{{\ttfamily 1911.11777}}].

\bibitem{IceCube:2021oiv}
{\scshape IceCube} collaboration, \emph{{Searching for High-Energy Neutrinos
  from Core-Collapse Supernovae with IceCube}},
  \href{https://doi.org/10.22323/1.395.1116}{\emph{PoS} {\bfseries ICRC2021}
  (2021) 1116} [\href{https://arxiv.org/abs/2107.09317}{{\ttfamily
  2107.09317}}].

\bibitem{2018PhDT.......322S}
A.~J. {Stasik}, \emph{{Search for high energetic neutrinos from core collapse
  supernovae using the IceCube neutrino telescope}}, Ph.D. thesis, Humboldt
  University of Berlin, Germany, Jan., 2018.

\bibitem{IceCube:2018omy}
{\scshape IceCube} collaboration, \emph{{Constraints on minute-scale transient
  astrophysical neutrino sources}},
  \href{https://doi.org/10.1103/PhysRevLett.122.051102}{\emph{Phys. Rev. Lett.}
  {\bfseries 122} (2019) 051102}
  [\href{https://arxiv.org/abs/1807.11492}{{\ttfamily 1807.11492}}].

\bibitem{Bartos:2021tok}
I.~Bartos, D.~Veske, M.~Kowalski, Z.~Marka and S.~Marka, \emph{{The IceCube Pie
  Chart: Relative Source Contributions to the Cosmic Neutrino Flux}},
  \href{https://doi.org/10.3847/1538-4357/ac1c7b}{\emph{Astrophys. J.}
  {\bfseries 921} (2021) 45}
  [\href{https://arxiv.org/abs/2105.03792}{{\ttfamily 2105.03792}}].

\bibitem{IceCube:2016ipa}
{\scshape IceCube} collaboration, \emph{{An All-Sky Search for Three Flavors of
  Neutrinos from Gamma-Ray Bursts with the IceCube Neutrino Observatory}},
  \href{https://doi.org/10.3847/0004-637X/824/2/115}{\emph{Astrophys. J.}
  {\bfseries 824} (2016) 115}
  [\href{https://arxiv.org/abs/1601.06484}{{\ttfamily 1601.06484}}].

\bibitem{Stein:2019ivm}
{\scshape IceCube} collaboration, \emph{{Search for Neutrinos from Populations
  of Optical Transients}},
  \href{https://doi.org/10.22323/1.358.1016}{\emph{PoS} {\bfseries ICRC2019}
  (2020) 1016} [\href{https://arxiv.org/abs/1908.08547}{{\ttfamily
  1908.08547}}].

\bibitem{ICECUBE:2021edr}
{\scshape ICECUBE} collaboration, \emph{{Search for High-Energy Neutrinos from
  Ultra-Luminous Infrared Galaxies with IceCube}}, {\emph{arXiv e-prints}
  (2021) } [\href{https://arxiv.org/abs/2107.03149}{{\ttfamily 2107.03149}}].

\bibitem{IceCube:2021abh}
{\scshape IceCube} collaboration, \emph{{A time-independent search for
  neutrinos from galaxy clusters with IceCube}},
  \href{https://doi.org/10.22323/1.395.1133}{\emph{PoS} {\bfseries ICRC2021}
  (2021) 1133} [\href{https://arxiv.org/abs/2107.10080}{{\ttfamily
  2107.10080}}].

\bibitem{2017ApJ...849...67A}
M.~G. {Aartsen}, M.~{Ackermann}, J.~{Adams}, J.~A. {Aguilar}, M.~{Ahlers},
  M.~{Ahrens} et~al., \emph{{Constraints on Galactic Neutrino Emission with
  Seven Years of IceCube Data}},
  \href{https://doi.org/10.3847/1538-4357/aa8dfb}{\emph{Astrophys. J.}
  {\bfseries 849} (2017) 67}
  [\href{https://arxiv.org/abs/1707.03416}{{\ttfamily 1707.03416}}].

\bibitem{IceCube:2021vlt}
{\scshape IceCube} collaboration, \emph{{Searching for time-dependent
  high-energy neutrino emission from X-ray binaries with IceCube}},
  \href{https://doi.org/10.22323/1.395.1136}{\emph{PoS} {\bfseries ICRC2021}
  (2021) 1136} [\href{https://arxiv.org/abs/2107.12383}{{\ttfamily
  2107.12383}}].

\bibitem{Zhou:2021rhl}
B.~Zhou, M.~Kamionkowski and Y.-f. Liang, \emph{{Search for High-Energy
  Neutrino Emission from Radio-Bright AGN}},
  \href{https://doi.org/10.1103/PhysRevD.103.123018}{\emph{Phys. Rev. D}
  {\bfseries 103} (2021) 123018}
  [\href{https://arxiv.org/abs/2103.12813}{{\ttfamily 2103.12813}}].

\bibitem{IceCube:2021pgw}
{\scshape IceCube} collaboration, \emph{{A search for neutrino emission from
  cores of Active Galactic Nuclei}},
  \href{https://arxiv.org/abs/2111.10169}{{\ttfamily 2111.10169}}.

\bibitem{Sasaki:2017uta}
{\scshape NTA} collaboration, \emph{{Neutrino Telescope Array (NTA):
  Multi-Astroparticle Explorer for PeV-EeV Universe\textemdash{} For Clear
  Identification of Cosmic Accelerators and Cosmic Beam Physics-}},
  \href{https://doi.org/10.22323/1.301.0941}{\emph{PoS} {\bfseries ICRC2017}
  (2018) 941}.

\bibitem{IceCube-Gen2:2020qha}
{\scshape IceCube-Gen2} collaboration, \emph{{IceCube-Gen2: the window to the
  extreme Universe}}, \href{https://doi.org/10.1088/1361-6471/abbd48}{\emph{J.
  Phys. G} {\bfseries 48} (2021) 060501}
  [\href{https://arxiv.org/abs/2008.04323}{{\ttfamily 2008.04323}}].

\bibitem{GRAND:2018iaj}
{\scshape GRAND} collaboration, \emph{{The Giant Radio Array for Neutrino
  Detection (GRAND): Science and Design}},
  \href{https://doi.org/10.1007/s11433-018-9385-7}{\emph{Sci. China Phys. Mech.
  Astron.} {\bfseries 63} (2020) 219501}
  [\href{https://arxiv.org/abs/1810.09994}{{\ttfamily 1810.09994}}].

\bibitem{POEMMA:2020ykm}
{\scshape POEMMA} collaboration, \emph{{The POEMMA (Probe of Extreme
  Multi-Messenger Astrophysics) observatory}},
  \href{https://doi.org/10.1088/1475-7516/2021/06/007}{\emph{JCAP} {\bfseries
  06} (2021) 007} [\href{https://arxiv.org/abs/2012.07945}{{\ttfamily
  2012.07945}}].

\bibitem{RNO-G:2020rmc}
{\scshape RNO-G} collaboration, \emph{{Design and Sensitivity of the Radio
  Neutrino Observatory in Greenland (RNO-G)}},
  \href{https://doi.org/10.1088/1748-0221/16/03/P03025}{\emph{JINST} {\bfseries
  16} (2021) P03025} [\href{https://arxiv.org/abs/2010.12279}{{\ttfamily
  2010.12279}}].

\bibitem{Brown:2021ane}
A.~M. Brown, M.~Bagheri, M.~Doro, E.~Gazda, D.~Kieda, C.~Lin et~al.,
  \emph{{Trinity: an imaging air Cherenkov telescope to search for
  Ultra-High-Energy neutrinos}},
  \href{https://doi.org/10.22323/1.395.1179}{\emph{PoS} {\bfseries ICRC2021}
  (2021) 1179}.

\bibitem{Neronov:2019htv}
A.~Neronov, \emph{{Sensitivity of top-of-the-mountain fluorescence telescope
  system for astrophysical neutrino flux above 10 PeV}},
  \href{https://doi.org/10.1016/j.astropartphys.2020.102549}{\emph{Astropart.
  Phys.} {\bfseries 128} (2021) 102549}
  [\href{https://arxiv.org/abs/1905.10606}{{\ttfamily 1905.10606}}].

\end{thebibliography}\endgroup
\end{multicols} 
\end{document}